\documentclass[12pt]{article}
\usepackage[compatibility=false]{caption}
\usepackage{cite}
\usepackage{amsmath,amssymb,amsfonts}
\usepackage{algorithmic}
\usepackage{graphicx}
\usepackage{textcomp}
\usepackage{amssymb}
\usepackage{lscape}
\usepackage{comment}
\usepackage{url}
\usepackage{amsmath}

\usepackage{graphicx}
\usepackage{caption}
\usepackage{comment}

\usepackage[utf8]{inputenc}        

\usepackage{authblk}  
\usepackage{xcolor}   
\usepackage{hyperref} 
\hypersetup{colorlinks=true, linkcolor=blue, urlcolor=blue, citecolor=blue}

\DeclareUnicodeCharacter{0080}{}   

\def\BibTeX{{\rm B\kern-.05em{\sc i\kern-.025em b}\kern-.08em
    T\kern-.1667em\lower.7ex\hbox{E}\kern-.125emX}}

\title{Simulation for All: A Step-by-Step Cookbook for Developing Human-Centered Multi-Agent Transportation Simulators}

\author[1]{Shiva Azimi}
\author[2]{Arash Tavakoli}

\affil[1]{Civil and Environmental Engineering, Villanova University, 800 E Lancaster Ave, Villanova, PA 19085, USA \\
Email: \textcolor{blue}{sazimi@villanova.edu}}
\affil[2]{Civil and Environmental Engineering, Villanova University, 800 E Lancaster Ave, Villanova, PA 19085, USA \\
Email: \textcolor{blue}{arash.tavakoli@villanova.edu}}

\date{}  

\begin{document}

\maketitle







\begin{abstract}
As cities evolve toward more complex and multimodal transportation systems, the need for human-centered multi-agent simulation tools has never been more urgent. Yet most existing platforms remain limited—they often separate different types of road users, rely on scripted or pre-defined behaviors, largely overlook public transit users as active, embodied participants, and are rarely designed with accessibility in mind for non-technical users. To address this gap, the following paper presents the detailed specifications of a multi-agent simulation platform designed to support real-time, human-centered, and immersive studies of all road users, accompanied by open-sourced scripts for replication. Using high fidelity immersive virtual environments, our platform enables interaction across public transit users, pedestrians, cyclists, automated vehicles, and drivers. The architecture is modular, extensible, and designed for accessibility. The system integrates hardware-specific modules—including an omnidirectional treadmill for pedestrians, a seating arrangement for public transit users, a smart trainer for cyclists, and an actuated cockpit for drivers. Additionally, the platform simultaneously collects multimodal physiological, neurological, and behavioral data through embedded sensing devices such as functional near-infrared spectroscopy (fNIRS) for brain activity, eye tracking, and wrist-based biosensors.  To show the usability of this system, we present three use cases across various areas of road user research. Simulation for All aims to pave the way for lowering the barrier to entry for high-fidelity transportation simulation, support experimentation across disciplines, and advance our understanding of multimodal mobility in increasingly complex urban environments. The codebase for the platform is available at \cite{ Azimi2025SimulationforAll}, enabling replication and adaptation for transportation research applications.
\end{abstract}



\maketitle

\section{Introduction}
Simulation has long been a vital tool in transportation research, offering a controlled environment to explore complex behaviors without the logistical, financial, and ethical challenges of real-world experiments \cite{dorokhin2020traffic,meuleners2015validation,papantoniou2015assessment,carsten2011driving,blana1996driving,huang2022overview,smith1995transims,cetin2002large,kotusevski2009review}. Historically, simulation platforms have supported a wide range of applications at both micro and macro levels \cite{dorokhin2020traffic}, from traffic flow modeling \cite{dorokhin2020traffic} to road user behavior analysis \cite{boyle2010using,taheri2017virtual,rosey2014driving,zhou2023car,meratexamining}, and more recently, the development of automated and semi-automated driving systems \cite{zangi2022driver,manawadu2015analysis,gonccalves2024using}. These platforms have provided researchers and industry stakeholders with powerful tools to test hypotheses, optimize system designs, and evaluate safety interventions \cite{caffo2020drives,bruck2020review}.

However, the origins of these platforms, reflect the historical development and disciplinary silos within transportation research. Simulators were typically developed for specific domains—such as driving safety, traffic engineering, or pedestrian dynamics with limited interdisciplinary integration \cite{bruck2020review,wynne2019systematic,boyle2010using}. Technical constraints, including limited computing power, and the absence of high-fidelity sensing technologies, further hindered the ability to simulate real-time, interactive behaviors among diverse road users \cite{miller2020flexible}. Simulation studies therefore often focused on individual modes of transportation \cite{sabeti2023mad}, initially drivers, then pedestrians \cite{schneider2020virtually,maillot2017training}, and more recently cyclists \cite{nazemi2021studying,ullmann2020bikevr,schulzyk2009real,guo2023psycho}—typically modeling each in isolation from others. 

As a result, behavioral variability shaped by culture, context, and individual differences was often reduced to deterministic or rule-based models due to computational or methodological limitations \cite{miller2020flexible}. While this approach has yielded valuable insights, recent research highlights its shortcomings in capturing the complex and interactive nature of real-world transportation systems, where diverse user groups coexist and influence each other's decisions \cite{lindner2022coupled,kalantari2023testing,sabeti2023mad,miller2020flexible}. 

Within the domain of road user behavior simulators, there is a growing push toward more realistic and interactive simulation environments particularly for studying multimodal interactions on the premise that simulations should reflect real-world conditions as closely as possible. This has become especially important with the current trend in automated vehicle testing and deployment. Multiple studies have emphasized the need to move beyond static or pre-programmed representations of "other" road users \cite{lindner2022coupled,kalantari2023testing,sabeti2023mad,miller2020flexible}. Studies show that generic or one-size-fits-all behavioral profiles fail to capture the variability and unpredictability that characterize real human interactions. These limitations are especially pronounced in urban environments, where pedestrians, cyclists, and drivers frequently interact at intersections, crosswalks, and shared spaces. Simulators that neglect these dynamics risk producing findings that do not generalize to real-world scenarios, hence lacking external validity \cite{wynne2019systematic}, especially in areas such as user trust, perceived safety, and behavioral compliance with traffic rules which is at the forefront of multiple user interaction.

Additionally, many of the current simulators often rely on proprietary software developed by commercial vendors \cite{sabeti2023mad}. As a result, they are typically closed systems, limiting access for new research groups who may lack the substantial financial resources or technical support required for customization. This reliance on proprietary solutions not only makes it difficult to add additional agent types—such as cyclists or micro-mobility users (e.g., scooterists), but also restricts the ability to modify core system functionalities, integrate new sensing modules, or develop novel behavioral models tailored to specific research needs. Moreover, proprietary simulators often come with rigid data structures and closed APIs, which can severely limit interoperability with emerging technologies such as AI-based behavior generation, physiological sensing devices, or new forms of environmental simulation (e.g., dynamic weather, smart infrastructure). 

Emerging VR platforms provide a compelling opportunity to model transportation interactions in the parallel multiplayer gaming environment fashion \cite{guo2023psycho,agudelo2021virtual,rebelo2012using,kuliga2015virtual,sabeti2023mad}. In games, multiple users operate in real time, make context-driven decisions, and respond dynamically to one another within immersive virtual worlds. Similarly, transportation simulators—particularly those focused on behavior—can benefit from these principles by enabling realistic, time-synchronized interactions among drivers, cyclists, and pedestrians. This shift from isolated, single-user experiences to shared, interactive environments mirrors the evolution seen in game design and opens up new possibilities for studying trust, conflict, negotiation, and coordination in transportation systems.

Coupled with VR, recent advancements in human sensing technologies have significantly expanded the potential of transportation simulations. Tools such as eye tracking, physiological monitoring (e.g., heart rate variability, skin conductance), and motion capture now enable researchers to capture nuanced indicators of attention, stress, cognitive workload, and emotional response in real time \cite{guo2023psycho,tavakoli2021leveraging,tavakoli2022driver,tavakoli2022multimodal}. When integrated with immersive VR environments, these sensing techniques provide a rich, multimodal understanding of how individuals experience and react to transportation scenarios, particularly during high-stakes or ambiguous situations.

While VR has had success in transportation research, the current landscape of transportation simulation is dominated by single-agent platforms and frameworks, with multi-agent systems only beginning to emerge in recent research over the past few years \cite{lehsing2017urban,kearney2018multi,bazilinskyy2020coupled,hubner2022external,lindner2022coupled,sabeti2023mad,lyu2024distributed,kwon2024investigating,feng2024does,fu2024immersive,crosato2024virtual,huang2025sky}. There remains a significant \textbf{lack of structured guidelines for designing, building, and extending multi-agent simulation environments.} This gap presents a major challenge for researchers aiming to support complex, real-time interactions among diverse road users. Additionally, the development and deployment of such platforms often demand advanced software engineering expertise, which may exceed the skillsets of many researchers—further limiting the adaptability and scalability of existing systems for new use cases and study designs.

Compounding this challenge is the rapidly growing landscape of human sensing technologies—including physiological monitors, eye trackers, and cognitive assessment tools—which are frequently difficult to integrate due to inconsistencies in data formats, hardware interfaces, and synchronization protocols. Moreover, while virtual reality has gained traction in transportation research, many current implementations are limited by physical space constraints, hindering scalability and restricting accessibility for broader academic and applied use.

The following paper aims to overcome the aforementioned challenges by presenting the development of the \textit{Human Centered Cities Lab Transportation Simulation Platform} \cite{humancenteredcities}, created at Villanova University. Rather than focusing on a single experiment or limited use case, this paper introduces a flexible and extensible simulation framework. \textbf{Our goal is to democratize the creation of multi-agent, human-centered transportation simulators by offering a step-by-step “cookbook” that is accessible to users from all technical backgrounds with minimal coding background. We provide detailed guidance, hardware specifications, and open-source scripts for data collection and analysis—empowering researchers, educators, and practitioners to build and adapt the system for their own studies.} 

More specifically, building on prior efforts in this space, this paper presents (1) a multi-agent simulation platform designed to support real-time, multimodal interaction among diverse transportation users—including, but not limited to, public transit users, automated vehicles, pedestrians, cyclists, and drivers—within a shared virtual environment; (2) this simulation system is integrated with a comprehensive human sensing module capable of capturing multiple dimensions of psychophysiology and well-being, including emotional responses, cognitive load, attentional focus, and physiological stress indicators; (3) in addition to detailed hardware implementation guidelines, we provide a customizable virtual city model developed using the Fantastic City Generator Unity asset \cite{fantastic_city_generator}, which can be adapted and extended by the research community for future experimental studies; (4) provide scripts for the system functionality.

The remainder of this paper is organized as follows: we first review the relevant background literature and existing simulation approaches together with their strengths and shortcomings in a detailed fashion. We then describe the methodology and technical architecture of our system. We subsequently present a series of use cases demonstrating the platform’s capabilities and conclude with a discussion of current challenges and directions for future development. We hope that this document allows more researchers to expand the horizon of development in this area to include other traffic agents beyond what is provided in this manuscript.

\section{Prior work in transportation simulation}
In the following sections, we review the state-of-the-art in transportation simulation, with a particular focus on the limitations and unmet needs within existing platforms—many of which are explored and addressed in our proposed methodology. It is important to recognize that these limitations often arose from historical technical constraints that were reasonable at the time, including restricted sensing capabilities and limited computational power. With recent advances in simulation technologies, human sensing, and real-time synchronization, there is now an opportunity to build upon prior efforts. \textbf{Rather than positioning this work as a replacement, our aim is to offer an updated, extensible platform that contributes to the broader research community and empowers diverse users to engage with human-centered, multi-agent transportation simulation.}

Broadly, transportation simulators can be categorized into two major types: system-level simulators and human-in-the-loop simulators. System-level simulators model traffic dynamics at the network or corridor scale, often using aggregated or rule-based representations of individual behavior. These tools are typically used for infrastructure planning, traffic management, and policy evaluation, such as VISSIM \cite{ptv2023vissim}, or SUMO \cite{krajzewicz2010traffic}. On the other hand, human-in-the-loop simulators, place human participants directly within the simulation environment, allowing researchers to capture detailed behavioral, cognitive, and emotional responses during dynamic interactions with the transportation system. These simulators are essential for studying decision-making, situational awareness, trust, compliance, and real-time human reactions under various traffic, environmental, and design scenarios \cite{wynne2019systematic,bruck2020review}.

Human-in-the-loop simulators also vary in terms of their fidelity, which refers to the degree to which the simulation replicates real-world conditions across sensory, cognitive, and behavioral dimensions \cite{kalantari2023testing,blana1996driving}. Fidelity encompasses multiple aspects, including visual realism, motion feedback, physical controls, auditory cues, and the psychological and cognitive demands placed on the user \cite{allen2007effect,himmels2024bigger}. High-fidelity simulators aim to create immersive, life-like experiences by closely mimicking the appearance, feel, and complexity of real-world driving, cycling, or walking environments \cite{greenberg2011physical,riener2010assessment}. These systems typically incorporate advanced visual displays, motion platforms, realistic vehicle cockpits or mobility devices, and synchronized environmental feedback such as sound, vibration, and resistance \cite{blana1996driving}. In contrast, low-fidelity simulators replicate only the essential components of a task, often using simplified visualizations and basic control interfaces \cite{blana1996driving}. Human-in-the-loop simulators can also be categorized as either single-agent or multi-agent systems, depending on the number of real users interacting within the simulated environment. 

\subsection{Single-agent simulators}
Single-agent simulators involve one user navigating a virtual space while other entities (e.g., vehicles, pedestrians) are controlled by scripted or AI-based agents. This includes driving, pedestrian, public transit, and bicycle simulators as follows.

\subsubsection{Driving Simulators}

Driving simulators are essential tools in automotive engineering and driving research, offering safe, controlled environments to study various aspects of driver behavior and human factors, evaluate vehicle and cockpit designs, examine roadway geometry and infrastructure features, test advanced driver assistance systems (ADAS) and autonomous technologies, and assess driver responses to varying road conditions, environmental factors, and in-vehicle distractions \cite{miller2020flexible,bruck2020review,boyle2010using,caffo2020drives,carsten2011driving}. Modern simulators vary widely in complexity, ranging from simple fixed-base desktop setups at the lower end of fidelity to sophisticated full-motion platforms that closely replicate the physical, visual, and auditory sensations of real-world driving \cite{bruck2020review}, available at multiple universities and research centers around the world. Many prior studies have detailed the development, testing, and validation of driving simulators (see \cite{freeman1995iowa,carlson1997iowa,khastgir2015development,ihemedu2015development,christoforou2025design} as examples); however, \textbf{it should be noted that these simulators often rely on proprietary software, which limits accessibility and constrains further development beyond the capabilities permitted by the underlying platform.}

\subsubsection{Pedestrian Simulators}
Pedestrian simulators generally consist of a virtual environment paired with motion controllers that allow participants to navigate the space using natural head and body movements \cite{tran2024evaluating,angulo2024evaluating,angulo2023demonstration}. At the high end of the fidelity spectrum for pedestrian simulators, CAVE (Cave Automatic Virtual Environment) systems project immersive virtual environments onto the walls and floor of a dedicated physical room, enabling participants to walk and explore without the need for a headset \cite{tabone2024immersive,zhu2025influences,pogmore2024virtual,yang2024using}. While CAVE systems offer high levels of immersion and spatial realism, they come at a significantly higher cost and require substantial infrastructural investment. Another type of pedestrian simulators uses VR type head mounted display (HMD) as the environment display, while the person walks in real physical space \cite{schneider2020virtually}. \textbf{Many of the prior VR based pedestrian simulators often require an extended amount of physical space for the participant to be able to walk within the environment.} This issue presents a gap in accessibility and spread of these research platforms across the community.

\subsubsection{Public Transit Simulators}

While public transportation plays a critical role in modern society, simulation-based studies have primarily focused on driver training rather than passenger experience. Most of the existing work in this domain centers on the development of training simulators for bus or rail operators \cite{forgefx_wmata_simulator}. However, there is a growing body of research and development aimed at simulating and improving the passenger experience—particularly in areas such as accessibility, comfort, and emotional responses to transit environments as follows.

Sadeghi et al. developed an immersive virtual subway simulation to examine how affective experiences and crowd density influence perceived travel time \cite{sadeghi2023affective}. Their findings show that increased virtual crowding reduced pleasantness and significantly lengthened subjective trip duration, with even small increases in density leading to measurable changes in time perception. Another study explored the applicability of VR for designing accessible public transportation services by engaging academic and industry experts in a series of co-design workshops \cite{burova2023virtual}. Burova et al. demonstrated that VR can support early-stage service design by enabling rapid testing and iteration of accessibility features in transit environments. The authors proposed a set of guidelines for integrating VR into the development of inclusive public transportation systems.

More recently, studies have investigated the potential of VR for supporting neurodiverse populations. Adjorlu et al. developed a VR simulation of a train station that enabled autistic adolescents to practice essential transit tasks such as reading schedules and navigating platforms, with results showing improvements in confidence and real-world attentiveness \cite{adjorlu2024virtual}. Similarly, Ravnsborg et al. designed a Unity-based VR bus training application that guided users through key travel actions, such as finding stops, checking in, and exiting at the correct location \cite{ravnsborg2025vr}.

Despite their contributions, these simulations remain limited in two important ways. First, \textbf{they are all single-agent systems—each involving only one real user in the simulation—thereby excluding the dynamic, multi-user interactions common in real-world transit environments. Second, they typically restrict users to a single transportation mode.} For example, a user cannot seamlessly transfer from walking to boarding a bus or train, a key element of multimodal travel. These constraints limit the ecological validity of such simulations and hinder their ability to capture the full spectrum of decisions, interactions, and transitions that occur in everyday public transportation use.

\subsubsection{Cyclist Simulators}

Cyclist simulators have emerged as valuable tools for studying cycling behavior, safety, and human factors within a controlled, risk-free environment. Based on the hardware and setup employed, cyclist simulators can be broadly categorized into stationary screen based or virtual reality based. Stationary bike setups utilize a physical bicycle or exercise bike in combination with a monitor or screen, enabling basic navigation through virtual environments \cite{sun2018design}. While cost-effective and easy to deploy, these setups typically offer limited immersion and rely on sensors mounted to the handlebars and pedals to track user input. In contrast, VR simulators significantly enhance immersion by integrating real-world cycling hardware—such as smart trainers or resistance-controlled stationary bikes—with virtual reality headsets \cite{guo2023psycho,wintersberger2022development,ullmann2020bikevr,matviienko2023does,hammami2025realistic}. This configuration allows users to physically pedal and steer while experiencing dynamic road environments, intersections, and interactions with other traffic agents. \textbf{Despite growing interest in electric bikes (e-bikes) as a critical component of urban micromobility, most existing simulators do not fully support e-bike-specific dynamics such as motor-assisted pedaling, acceleration curves, or battery management.} This presents a gap in the current simulation landscape, as e-bikes exhibit distinct behavioral and performance characteristics that influence rider decision-making, safety margins, and interaction with other road users. Incorporating e-bike features into cycling simulators is essential for studying their role in traffic systems, particularly as cities adapt to the rising popularity of electrically assisted micromobility.

\subsection{Multi-agent Simulators}
While single-agent simulators are valuable for studying individual behavior, this approach overlooks the complex, dynamic interactions that occur between multiple road users in real-world transportation settings limiting the ecological validity of findings. In response to these limitations, recent research efforts have aimed to incorporate multiple agents within a shared simulation environment, enabling more realistic and dynamic studies of multimodal interactions. 

We conducted a comprehensive literature review of multi-agent virtual environment studies. Table \ref{tab:multi} summarizes existing multi-agent, human-in-the-loop simulation platforms and highlights their respective capabilities in human sensing. \textbf{Notably, the majority of these simulators include only two real agents, limiting the ability to study more complex, multimodal interactions. Furthermore, none of the reviewed studies incorporate public transit users—despite their critical role in real-world transportation systems.}


\begin{table*}[htbp]
\centering
\resizebox{\textwidth}{!}{%
\begin{tabular}{|p{2cm}|p{1cm}|p{1.3cm}|p{2cm}|p{1.3cm}|p{2cm}|p{1.3cm}|p{2cm}|p{5cm}|p{2.7cm}|p{2.7cm}|}

\hline
\multicolumn{1}{|c|}{\textbf{Reference}} & 
\multicolumn{1}{c|}{\textbf{\#Ag}} & 
\multicolumn{1}{c|}{\textbf{Agent 1}} & 
\multicolumn{1}{c|}{\textbf{Platform 1}} & 
\multicolumn{1}{c|}{\textbf{Agent 2}} & 
\multicolumn{1}{c|}{\textbf{Platform 2}} & 
\multicolumn{1}{c|}{\textbf{Agent 3}} & 
\multicolumn{1}{c|}{\textbf{Platform 3}} & 
\multicolumn{1}{c|}{\textbf{Behavioral}} & 
\multicolumn{1}{c|}{\textbf{Physiological}} & 
\multicolumn{1}{c|}{\textbf{Psychological}} \\
\hline

\cite{lehsing2017urban} & 2 & Driver & Full cabin driving simulator & Ped & CAVE & - & - & Speed, Acceleration, Braking pressure, Trajectory, Crossing decision, Yielding behavior, Time to Arrival, Deceleration-to-Safety Time, Time to React, Behavioral synchronization & - & - \\
\hline
\cite{kearney2018multi} & 2 & Driver & Full cabin driver simulator & Ped & VR & - & - & Ped crossing behavior, Driver yielding behavior, Vehicle following distance, Teleportation response, Agent trajectory behavior & - & - \\
\hline
\cite{perez2019ar} & 2 & Ped & AR (Microsoft HoloLens) & Driver & Desktop Simulator & - & - & Pedestrian position, Walking path, Crossing time, Real-world path comparison, Multi-user interaction with driver-controlled vehicles & - & Realism of movement assessed via behavior comparison and post-experiment analysis \\
\hline
\cite{bazilinskyy2020coupled} & 3 & AV & Desktop Driving Simulator & Driver & Desktop Driving Simulator & Ped & VR & Vehicle speed, Braking input, Steering input, eHMI activation, Ped crossing decision, Motion activity factor, Elbow motion & - & - \\
\hline
\cite{hubner2022external} & 2 & Driver & Cockpit Driving Simulator & Ped & VR & - & - & Crossing initiation time, Crossing time, Passing time, Minimum Time to Collision, Subjective error rate, Objective error rate & - & - \\
\hline
\cite{lindner2022coupled} & 2 & AV & Desktop Simulator & Ped & VR & - & - & Speed, Steering angle, Braking behavior, Hand signals, Conflict resolution decisions & - & Assessed via post-experiment perceived safety surveys \\
\hline
\cite{sabeti2023mad} & 3 & Driver & Desktop Simulator & Ped & VR & Cyclist & VR & Crossing decision, Bicyclist reaction time, Steering input, Braking behavior, Driver yielding behavior, Multi-agent interaction & - & - \\
\hline
\cite{lyu2024distributed} & 2 & Driver & Desktop Driving Simulator & Ped & CAVE & - & - & Crossing decision, Head-turning frequency, Head-turning rate & Eye Tracking: Used to analyze gaze direction, fixation on vehicles, and attentional focus & - \\
\hline
\cite{kwon2024investigating} & 2 & Ped & VR & Ped & VR & - & - & Walking speed, Route choice, Proximity to vehicles, Waiting time, Crossing decision, Visual fixation patterns & Eye Tracking was used to measure fixation and visual attention & - \\
\hline
\cite{feng2024does} & 2 & Ped & VR & Ped & VR & - & - & Crossing initiation time, Time before crossing, Time to cross, Vehicle-gazing time, Crossing speed, Total crossing distance, Space gap & Eye Tracking was used to measure visual attention, particularly vehicle-gazing & Perceived risk \\
\hline
\cite{fu2024immersive} & 2 & Ped & VR & Ped & VR & - & - & Speed, Acceleration, Direction, Hand gestures, Light/horn use, Yielding behavior & - & - \\
\hline
\cite{crosato2024virtual} & 2 & Driver & Cockpit Simulator & Ped & VR & - & - & Crossing decision, Time to Collision, Ped trajectory, Vehicle trajectory, Driver braking, Ped pose & - & - \\
\hline
\cite{huang2025sky} & 3 & AV & Desktop Simulator & Driver & Desktop Simulator & Ped & VR & Gaze patterns, Voice commands, Facial expressions, Steering input, Pedal input, Interaction with AV agents & Heart rate and HRV measured via smartwatch; Via HTC Vive Pro Eye & Stress and affect via facial expressions and voice, cognitive load inferred through task complexity and system interaction \\
\hline
\end{tabular}
}
\caption{Summary of multi-agent simulation studies with their behavioral, physiological, and psychological measures}
\label{tab:multi}
\end{table*}

\subsection{Human Sensing}

As simulation platforms have evolved to more accurately replicate real-world conditions, human sensing has emerged as a critical tool for gaining deeper insights into how individuals experience and respond to transportation environments. Beyond observing external behavior, human sensing allows researchers to capture a wide range of internal states—physiological, psychological, and behavioral—that influence decision-making, attention, stress, and safety outcomes.

Human sensing within simulators can be broadly categorized into three dimensions:

\subsubsection{Physiological Sensing}
Physiological measurements provide real-time indicators of the body's responses to external stimuli. Common physiological sensing modalities include cardiovascular monitoring (e.g., heart rate, heart rate variability), neuroimaging techniques such as functional near-infrared spectroscopy (fNIRS) or electroencephalography (EEG) for assessing brain activity, and ocular tracking (eye tracking) to capture gaze behavior, pupil dilation, and attentional focus \cite{shiferaw2019gaze,bunce2006functional,tavakoli2022driver}. Additionally, skin conductance (electrodermal activity) and skin temperature measurements are often used as proxies for emotional arousal and stress responses \cite{guo2024unveiling,izzetoglu2020short,gomero2022evaluation}. 

\textbf{While physiological sensing has become a central component of modern simulation research, many existing platforms offer only limited capabilities—often restricted to basic wearable devices such as smartwatches and short-term eye tracking using third-party software on VR goggles.} While these technologies have improved in recent years, research suggests they are often less accurate and less robust than specialized tools such as headband-style neuroimaging sensors or clinical-grade ECG systems. As a result, the granularity and reliability of physiological data in prior simulation studies remain limited. More advanced modalities, including electroencephalography (EEG) and functional near-infrared spectroscopy (fNIRS), remain underutilized despite their potential to provide deeper insights into cognitive and neural states. One contributing factor has been the physical incompatibility between head-mounted VR displays and the headband-style sensors required for brain activity monitoring. However, recent developments in lightweight and compact fNIRS headbands now allow for simultaneous data collection alongside VR goggles, enabling a more integrated and comprehensive sensing approach.

In parallel, VR headset technology has significantly advanced, particularly in embedded eye-tracking capabilities. Earlier systems such as the HTC Vive Pro Eye typically support sampling rates of 90–120 Hz and often rely on external software development kits (SDKs) for data extraction and processing. In contrast, newer headsets such as the Varjo XR-4 offer integrated eye tracking at 200 Hz, enabling higher temporal resolution without the need for third-party software. \textbf{This improvement enhances the precision of gaze data—supporting more accurate analyses of visual attention, cognitive load, and user intent in complex, interactive environments.}

\subsubsection{Psychological Sensing}
Psychological assessments typically involve subjective measures administered through questionnaires and standardized scales. These tools capture perceived mental workload, stress levels, affective states, trust, risk perception, and situational awareness. Commonly used instruments include the NASA Task Load Index (NASA-TLX) for cognitive workload \cite{hart1988development}, the Positive and Negative Affect Schedule (PANAS) for emotional states \cite{mauss2009measures}, and domain-specific stress perception \cite{karvounides2016three}. \textbf{Psychological assessments in prior studies are frequently administered outside the simulation environment, typically via separate survey platforms. This disjointed approach increases the risk of disrupting user immersion and breaking the sense of presence—both of which are critical for maintaining ecological validity in virtual simulations \cite{safikhani2021influence}.} To advance human-centered simulation research, there is a growing need for more seamless integration of physiological and psychological sensing within the simulation platforms themselves.

\subsubsection{Behavioral Sensing}
Behavioral sensing involves the direct measurement of user actions and performance within the simulation environment. In driving simulators, this may include metrics such as lane-keeping behavior, braking reaction times, gap acceptance, speed variability, and collision avoidance maneuvers. In cycling simulators, behavioral data often capture steering stability, pedal cadence, evasive actions, and interactions with virtual traffic. In pedestrian simulators, walking speed, path trajectories, gap crossing decisions, and stop-and-go patterns are commonly monitored. In the context of public transit simulations, behavioral metrics may include boarding and alighting times, dwell time at stations, seat selection behavior, proximity to other passengers, reaction to crowding, and navigation through stations and cabins. These actions can be particularly informative in understanding accessibility, comfort, and wayfinding in multimodal journeys. Across all modes, these behavioral outputs are crucial for linking observed actions to underlying physiological and psychological states.

\section{Methodology}
In the following subsections, we detail the design and implementation of our multimodal, human-centered simulation platform. This includes the overall system architecture, hardware configurations for each agent type (pedestrian, cyclist, and driver), integration of human sensing technologies, and tools for scenario development and synchronization. A complete list of hardware components required to replicate or extend the system is provided in Appendix~\ref{appendix:hardware}. Implementation details, setup instructions, and source code are available at the project’s GitHub repository~\cite{Azimi2025SimulationforAll}.

\subsection{System Architecture}

This section focuses on the overall system architecture that enables real-time, human-in-the-loop interaction among multiple road users. While detailed descriptions of the individual agent hardware and sensing modules are provided in subsequent sections, here we concentrate on the system-level design and integration strategy that underpins the entire simulation platform.

\begin{figure}[htbp]
    \centering
    \includegraphics[width=1\textwidth]{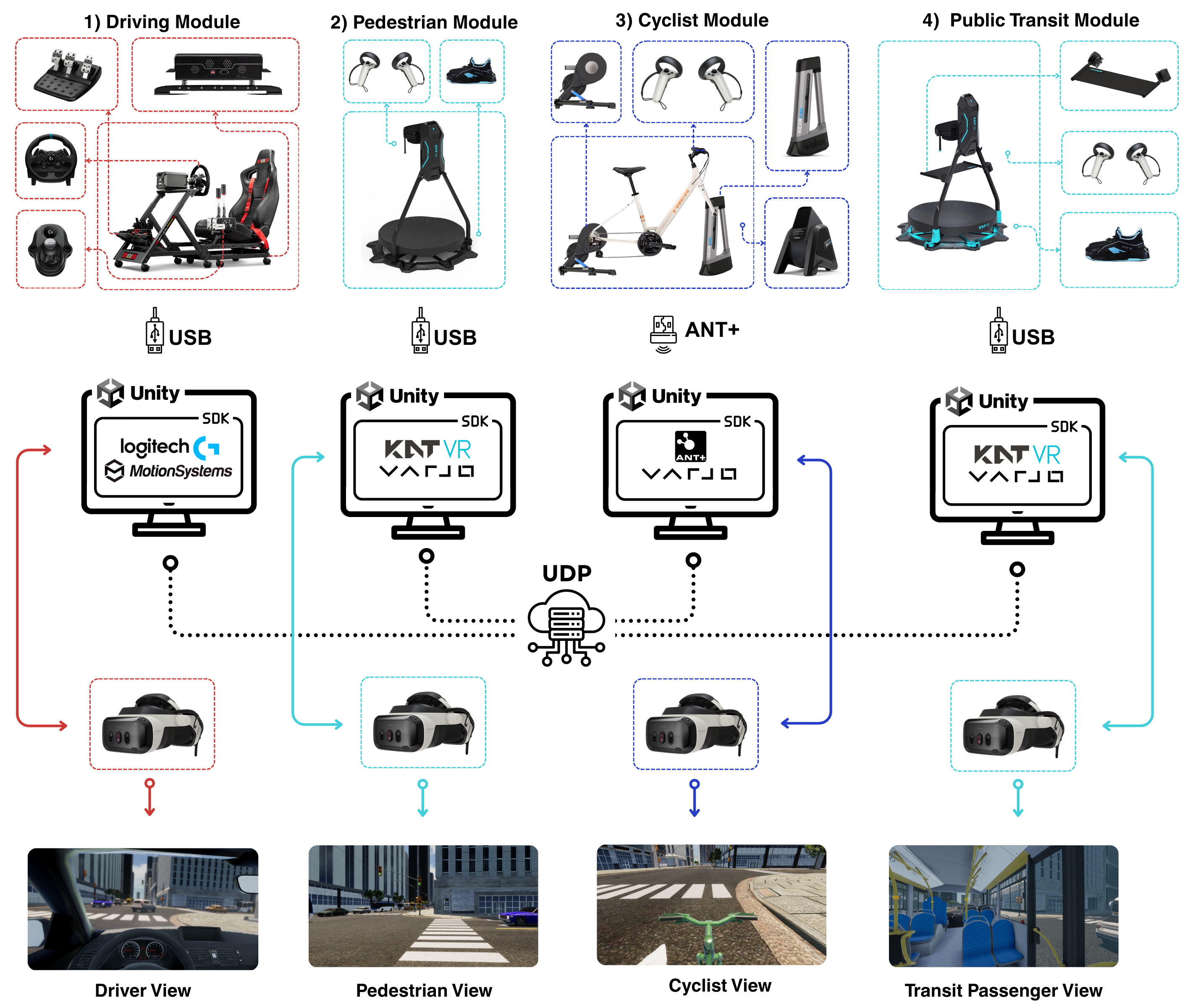}
    \captionsetup{justification=centering}
    \caption{System architecture outlining the components and connections for each agent in the simulation—driver, pedestrian, cyclist, and public transit user. Each agent is equipped with a dedicated computing engine and integrated with its corresponding physical interface}
    \label{fig:system-architecture}
\end{figure}

Figure~\ref{fig:system-architecture} illustrates the architecture of our immersive, modular, multi-agent simulation environment. The system includes various agent-specific hardware components that are essential for interaction within the simulation. Each user—whether a cyclist, pedestrian, or driver—operates through distinct physical systems, such as a cycling trainer, omnidirectional treadmill, or driving cockpit. All hardware interfaces with the simulation environment through dedicated software development kits (SDKs) and standard communication protocols such as USB or Bluetooth. This modular approach ensures flexibility, extensibility, and real-time coordination across different agents in the virtual environment.

Next, the system includes a network of computing engines assigned to each agent. These computers are synchronized using a low-latency communication protocol based on User Datagram Protocol (UDP), ensuring real-time data exchange and temporal alignment across all agents in the simulation. To maintain high-fidelity interactions and scalability, the number of computing engines must match the number of active agents, such as one for the pedestrian, cyclist, and driver modules.

Finally, the sensor data from each agent is transmitted into a shared virtual environment built in Unity. This environment acts as the interface between the physical hardware and the digital simulation. Real-time sensor readings—such as steering angle, pedaling speed, or walking direction—are used to drive each agent’s behavior within Unity. For example, when a driver turns the steering wheel, the vehicle model in Unity responds by rotating. Similarly, a cyclist’s pedaling intensity controls speed, and a pedestrian’s walking direction guides movement through the scene. This integration ensures that each user’s physical actions are immediately reflected in the virtual world, enabling high-fidelity interaction among all agents.

\subsubsection{Driver Module}
\begin{figure}[htbp]
    \centering
    \includegraphics[width=0.5\textwidth]{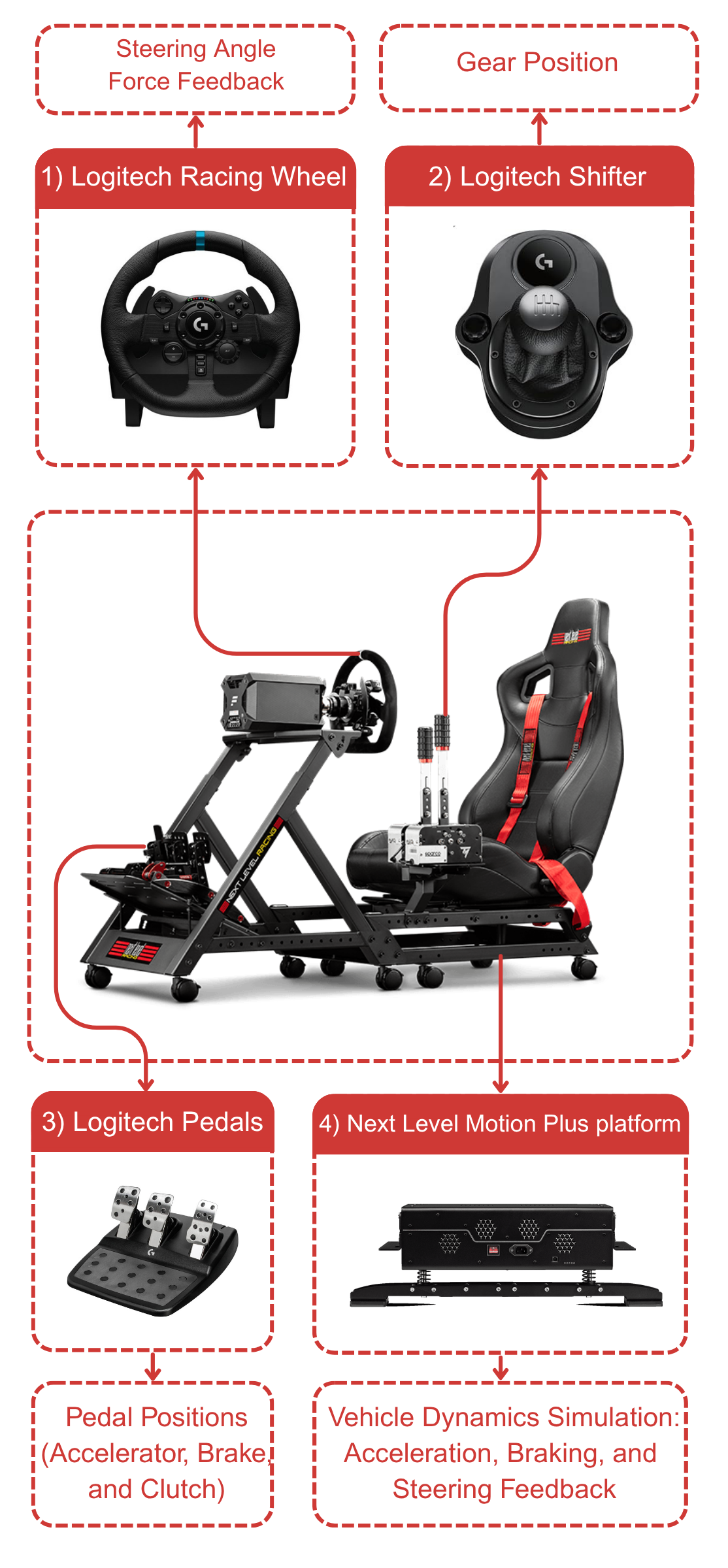} 
    \captionsetup{justification=centering}
    \caption{Driving Module Hardware and Data Collection: 1) Logitech Racing Wheel, 2) Logitech Shifter, 3) Logitech Pedals, and 4) Next Level Motion Plus Platform}
    \label{fig:driver-module}
\end{figure}

The Driving Simulator integrates real-world hardware with virtual reality (VR) to simulate realistic driving scenarios. As shown in Figure~\ref{fig:driver-module}, the setup includes the GT-Track Simulator Cockpit, Next Level Motion Plus, and G923 Racing Wheel, Pedals, and Shifter, which work together to provide an immersive and interactive driving experience.\\

Physical Equipment and Sensors of the driving module is as follows:
\begin{itemize}
    \item \textbf{GTTrack Simulator Cockpit} \\
    The GTTrack Simulator Cockpit serves as the primary vehicle chassis for the driving simulator. It provides a realistic seating position and adjustable controls for the G923 Racing Wheel and Next Level Motion Plus, ensuring a dynamic and immersive experience.\\

    \item \textbf{Next Level Motion Plus Platform} \\
    The Next Level Motion Plus platform simulates force feedback and vehicle dynamics by adjusting the cockpit position in real-time, enhancing the realism of acceleration, braking, and steering.\\

    \item \textbf{G923 Racing Wheel, Pedals, and Shifter} \\
    The Logitech G923 Racing Wheel controls the direction of the vehicle, the pedals control braking and acceleration, and the shifter manages gear changes. Each component provides force feedback, with the wheel offering resistance and vibrations to simulate road conditions, the pedals providing resistance for braking and acceleration, and the shifter giving tactile feedback during gear transitions.

\end{itemize}

\subsubsection{Automated Vehicle (AV) Module}
The Automated Vehicle (AV) module is designed to simulate real-world autonomous driving behavior while allowing for optional human supervision, similar to advanced driver assistance systems found in commercial vehicles such as Tesla’s Autopilot. In this configuration, the AV operates in a fully automated mode using pre-defined behavioral scripts or AI-driven navigation within the Unity environment, while a human user seated in the simulator can observe its behavior and intervene if necessary. This setup allows researchers to study trust, perception, and supervisory behaviors associated with semi-autonomous systems.

Physical Equipment and Sensors of the AV module are as follows:

\begin{itemize}
\item \textbf{GTTrack Simulator Cockpit and Display Setup} \\
The AV shares the same physical setup as the human driver, including the GTTrack cockpit, a steering wheel, and pedals. The system continuously logs automated driving dynamics, such as AV speed and acceleration profiles, to characterize its behavior. However, during the AV scenario, these controls are not actively used unless a user is prompted to take over. A large forward-facing display or VR headset presents the simulated environment from the vehicle’s perspective.
\item \textbf{Automated Navigation Logic (Unity-based)} \\
The AV agent navigates the environment using a custom Unity script that integrates sensor-like data (e.g., lane position, obstacle distance, traffic light state) to simulate automated driving behavior. Navigation decisions such as lane-keeping, acceleration, and yielding are governed by state-machine logic or reinforcement learning–based policy models, while the system concurrently computes time-to-collision and vehicle-proximity metrics to quantify safety margins in real time.

\item \textbf{Supervisory Interface and Takeover Control} \\
A manual override mechanism is included, allowing the user to take control of the vehicle via steering wheel, pedals, and shifter. This is triggered in scenarios that simulate edge cases or system faults (e.g., blocked lane, pedestrian dart-out), enabling the study of takeover response time, situational awareness, and trust calibration.

\item \textbf{Eye Tracking and Physiological Sensors} \\
When a user is present in the AV module, the system records gaze behavior and physiological states to assess supervisory attention, stress, and cognitive load. Eye tracking data is captured via the Varjo VR headset to determine whether users monitor the road or become disengaged. Simultaneous physiological monitoring (e.g., HRV, EDA) provides insights into the user’s readiness to intervene.

\end{itemize}

This AV module supports both fully autonomous operation and human-in-the-loop supervision, making it a flexible tool for studying human trust, system transparency, and safety behaviors in semi-automated driving scenarios.

\subsubsection{Pedestrian Module}
This section outlines the components of the Pedestrian Simulator, including the hardware setup, the sensors used, and the data collected to track the pedestrian’s behavioral responses in the virtual environment. \\

\begin{figure}[htbp]
    \centering
    \includegraphics[width=0.5\textwidth]{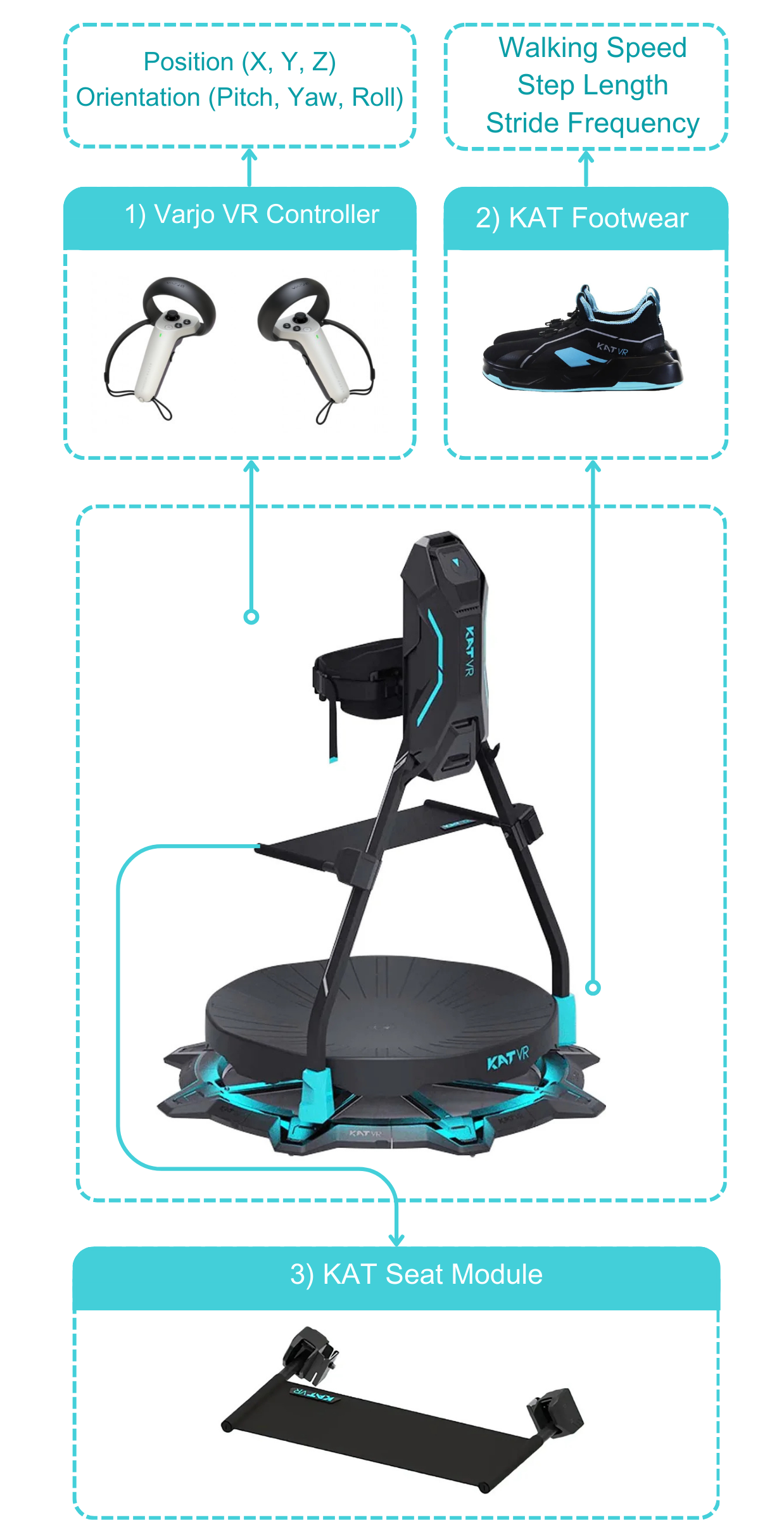} 
    \captionsetup{justification=centering}
    \caption{Pedestrian and Public Transit Module Hardware and Data Collection: 1) Varjo VR Controller, 2) KAT Footwear, with the addition of 3) KAT Seat Module in the Public Transit Module}
    \label{fig:pedestrian-module}
\end{figure}

As shown in Figure~\ref{fig:pedestrian-module}, the physical equipment and sensors of the pedestrian module are as follows:

\begin{itemize}
    \item \textbf{Kat Walk VR Core 2+} \\
The Kat Walk VR Core 2+ omnidirectional treadmill is designed to facilitate realistic pedestrian movement within a virtual environment, allowing participants to engage in free walking in all directions and providing an immersive simulation of real-world behavior. This system was selected for its ability to offer 360-degree walking freedom and real-time movement tracking, and it records walking speed (in meters per second), step length, and turning angles as the participant navigates through the virtual space.

 \item \textbf{Back Support System} \\
The Kat Walk VR Core 2+ is equipped with a vehicle-mounted back support system that is attached to the participant's wrist and back. This system tracks the participant's orientation and direction of movement in real time. By capturing the spinal orientation, the system allows for accurate simulation of the pedestrian's walking trajectory and path adjustments when interacting with other road users. This feature provides more precise data than traditional motion-capture systems and ensures accurate tracking of the pedestrian’s walking behavior.

\end{itemize}

\subsubsection{Public Transit Module}

This section outlines the components of the Public Transit Simulator, including the hardware setup and system configuration used to simulate realistic movement through transit environments such as stations, platforms, and vehicle cabins. The simulator is designed to allow participants to experience boarding, riding, and exiting a bus or train within a continuous immersive scenario. Figure~\ref{fig:pedestrian-module} illustrates the hardware setup, which includes components shared with the pedestrian module, as well as additional equipment specific to the public transit simulation.

Physical Equipment of the public transit module are as follows:

\begin{itemize}
    \item \textbf{VARJO XR-4 Headset} \\
The VARJO XR-4 headset is used to deliver a high-fidelity, immersive visual experience that includes both station and in-vehicle environments. It features integrated positional tracking and enables the user to engage with dynamic transit scenarios, such as navigating a crowded cabin or reading informational signage. The system continuously tracks head movement and viewing direction to reflect the user’s orientation and navigation within the simulation.
    
    \item \textbf{Kat Walk VR Core 2+} \\
The Kat Walk VR Core 2+ omnidirectional treadmill allows the participant to walk through virtual transit spaces—such as platforms, terminals, and aisles—without requiring a large physical footprint. It supports free movement in all directions and enables seamless transitions between walking and stationary states, and the system tracks walking speed, direction, and turning behavior as users move through stations and toward boarding areas.
    
    \item \textbf{Seated Mode Configuration} \\
    During the in-cabin portion of the simulation, participants can transition from walking to sitting to simulate a realistic ride experience. The system supports this mode by allowing the user to remain in the same virtual environment, preserving spatial continuity from station entry to disembarkation.
\end{itemize}

\subsubsection{Cyclist Module}

\begin{figure}[htbp]
    \centering
    \includegraphics[width=0.5\textwidth]{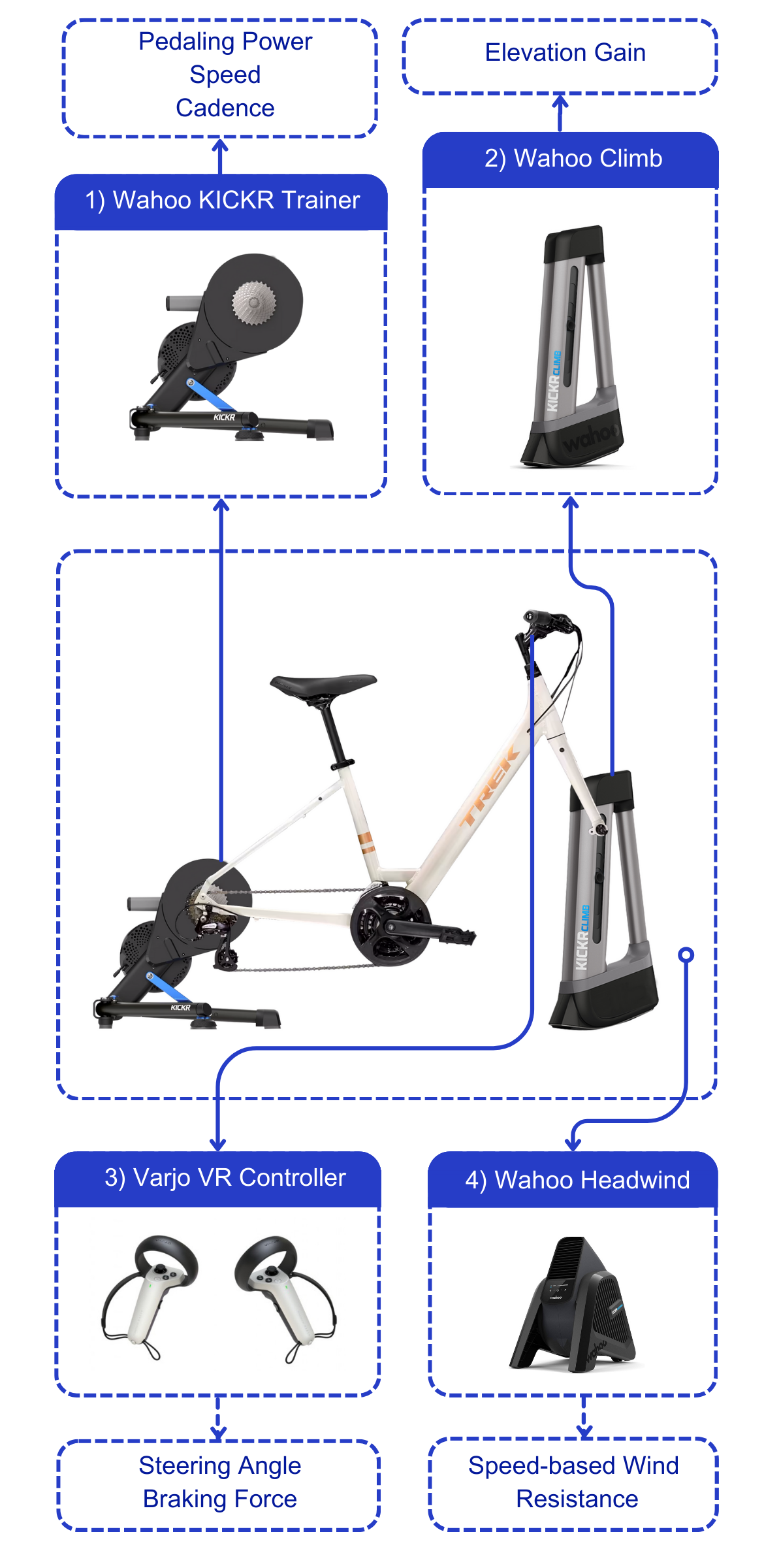} 
    \captionsetup{justification=centering}
    \caption{Cyclist Simulator Hardware and Data Collection: 1) Wahoo KICKR Trainer, 2) Wahoo Climb, 3) Varjo VR Controller, and 4) Wahoo Headwind}
    \label{fig:cyclist-module}
\end{figure}

Building on the work provided in \cite{guo2023psycho}, the cyclist module integrates real-world cycling hardware with sensor technologies and a VR environment to simulate immersive and interactive cycling behavior. The system enables the cyclist to operate within a shared traffic scenario, interacting in real time with both pedestrian and vehicular agents controlled by real users. However, prior work lacks more recent implementation of cycling, including e-bike situations. In contrast to prior works in this area, this system is also capable of simulating electrically assisted bicycles (e-bikes). This is achieved through integrated control logic that adjusts resistance and motion parameters based on pedal-assist profiles, allowing for the exploration of how e-bike dynamics influence user behavior and interaction in shared traffic environments. Figure~\ref{fig:cyclist-module} shows the hardware components of the cyclist module. The cyclist module is fully networked with the pedestrian and driver simulators via a low-latency protocol, allowing for seamless and time-synchronized interactions. 

Physical Equipment and Sensors of the Cycling Module are as follows:

\begin{itemize}
    \item \textbf{Wahoo KICKR Trainer} \\
    The Wahoo KICKR Trainer is used as the primary resistance unit to simulate cycling efforts in the virtual environment. It mounts a Trek Verve 1 bike and provides real-time feedback on the cyclist’s performance. The system collects pedaling power, speed, and cadence: pedaling power measures the force exerted by the rider to maintain speed and navigate virtual terrain; speed tracks how fast the cyclist moves, adjusting resistance for uphill or downhill conditions; and cadence monitors revolutions per minute to capture the cyclist’s rhythm and effort throughout the ride.

    \item \textbf{Wahoo Climb} \\
    The Wahoo Climb simulates road elevation by adjusting the angle of the bike to reflect changes in the virtual terrain. The system collects elevation gain data, measuring the virtual road’s incline and automatically tilting the bike to replicate uphill and downhill sections of the ride.

    \item \textbf{Wahoo Headwind} \\
    The Wahoo Headwind is integrated into the system to simulate real-world wind resistance based on the cyclist’s speed and the terrain, adjusting fan intensity dynamically: it increases wind output as the cyclist’s speed rises to replicate the airflow experienced at higher speeds, and during downhill simulation it generates stronger airflow to mirror the wind’s effect when descending virtual hills.

    \item \textbf{The Varjo XR-4 VR headset and controller} \\
The Varjo XR-4 VR headset and controller are used to create a realistic cycling simulation within the shared VR environment, immersing the user in dynamic visual and environmental feedback. The headset’s eye-tracking sensors capture the cyclist’s gaze direction and visual attention to analyze how stimuli—such as pedestrians crossing the road or oncoming vehicles—impact decision-making and reaction times. The controller tracks hand movements to measure steering angle and braking force, enabling real-time interaction with the virtual terrain and other agents.
\end{itemize}


\subsection{Human Sensing Module}

\begin{figure*}[htbp]
    \centering
    \includegraphics[width=\textwidth]{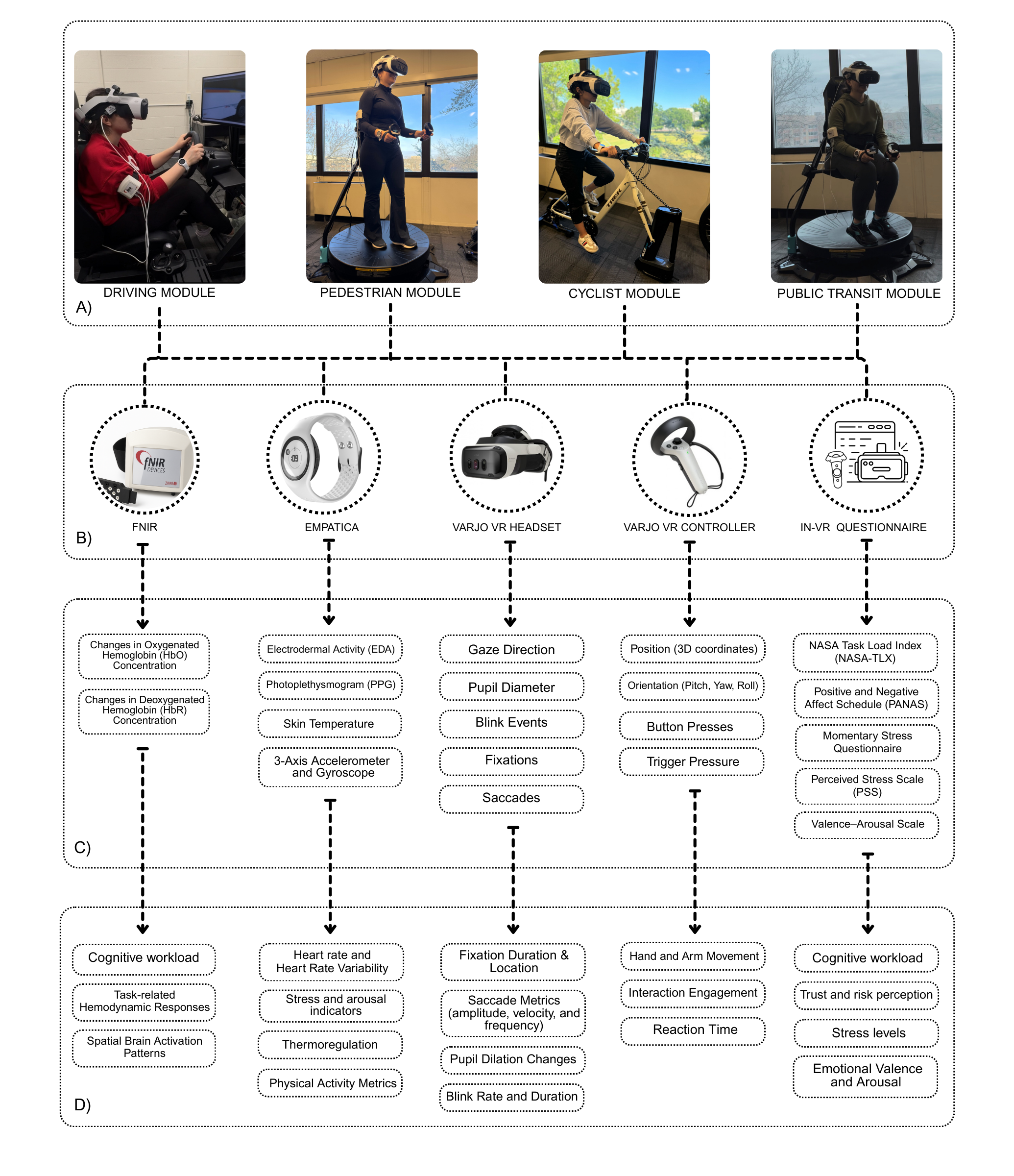} 
    \captionsetup{justification=centering}
    \caption{ Human Sensing Module A) Modules, B) Devices, C) Data Collection, D) Feature Extraction}
    \label{fig:human-sensing}
\end{figure*}

To enable comprehensive analysis of cognitive, emotional, and physiological responses during transportation interactions, our simulation platform integrates a dedicated human sensing module. This module collects multimodal data from all participating agents (cyclist, pedestrian, and driver/passenger) using wearable and embedded sensors. The following sensing mechanisms are currently supported in the system and are shown in Figure~\ref{fig:human-sensing}. 

\begin{figure}[htbp]
    \centering
    \includegraphics[width=1\textwidth]{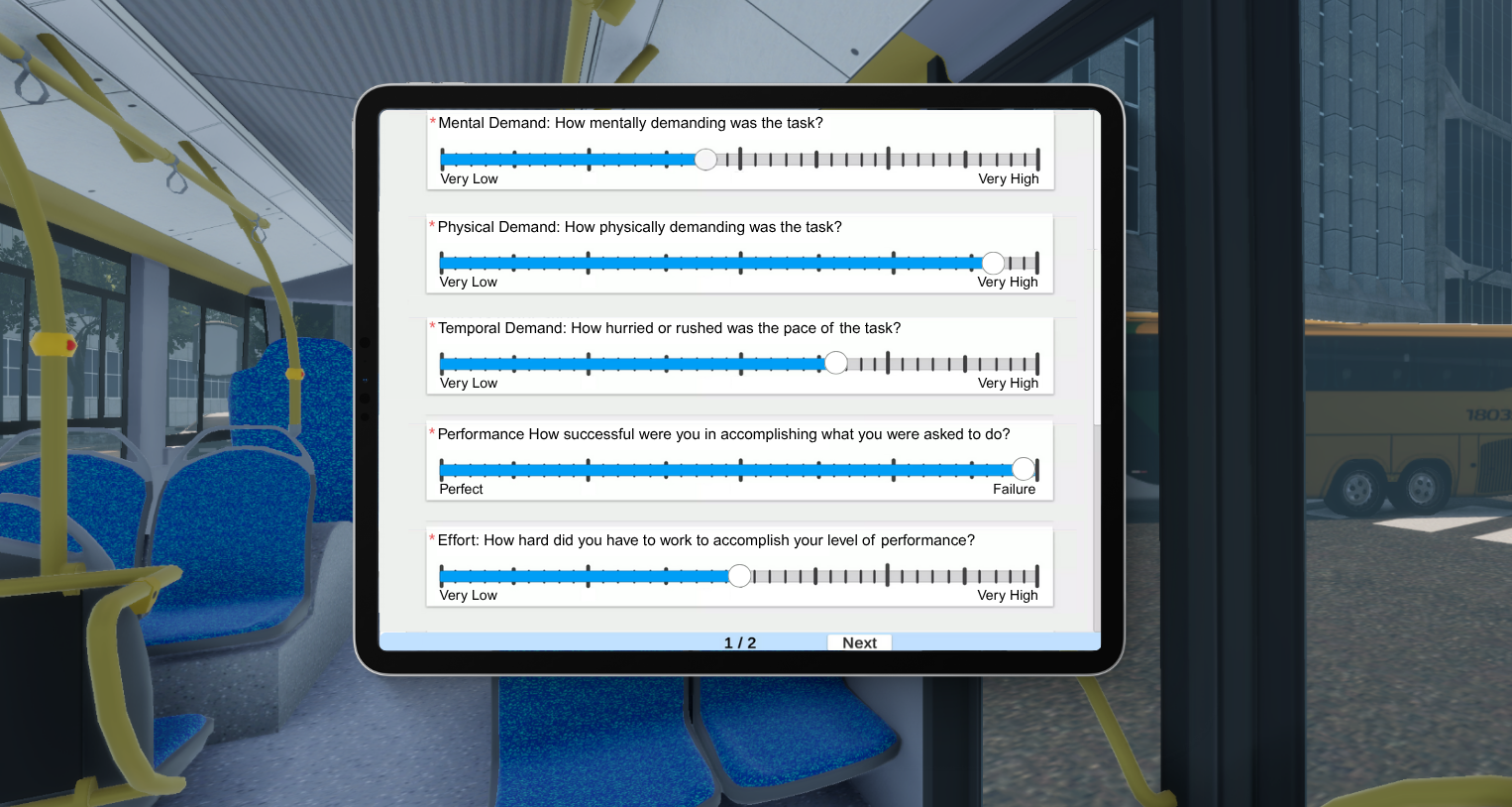} 
    \captionsetup{justification=centering}
    \caption{In-VR questionnaire interface for collecting psychological responses during immersive transportation simulation}
    \label{fig:invr-questionnaire}
\end{figure}

Additionally, the module collects subjective and objective standard scale data within the simulation using an in-VR questionnaire, as shown in Figure~\ref{fig:invr-questionnaire}. The combination of these sensing techniques provides a rich dataset for evaluating user state in real time, including stress, emotion, cognitive workload, and attention.

\subsubsection{Physiological Sensors}
\begin{itemize}
    \item \textbf{Empatica EmbracePlus}: This wearable wristband provides continuous measurement of physiological indicators including electrodermal activity (EDA), heart rate (HR), heart rate variability (HRV), skin temperature, and accelerometry. These metrics serve as proxies for emotional arousal, stress response, thermoregulation, and physical movement, respectively. The EmbracePlus is particularly valuable for assessing acute stress during high-conflict or ambiguous interactions (e.g., near-misses or sudden events).

    \item \textbf{fNIRS Headband}: Functional near-infrared spectroscopy (fNIRS) allows for non-invasive monitoring of cerebral hemodynamics by measuring changes in oxygenated and deoxygenated hemoglobin. Our system uses a headband-style fNIRS sensor to assess cognitive workload and attentional resource allocation. This sensor is especially useful in scenarios involving decision-making under uncertainty, multitasking, or human-AV interactions that demand sustained vigilance.

    \item \textbf{Eye Tracker (Integrated in VR Headsets)}: Eye tracking is embedded in the Varjo VR headset and is used to monitor gaze direction, fixation duration, and saccadic patterns. This data helps identify what participants attend to in the environment, how visual information guides decision-making, and whether agents respond to social cues such as eye contact or visual motion from other users. Eye tracking also enables the creation of gaze heatmaps for post-scenario analysis.

\end{itemize}

All sensing streams are time-synchronized with simulator events via software-level integration, allowing for detailed alignment between human state data and key interaction points. This multimodal sensing setup enhances the ecological validity of behavioral analysis and supports both moment-to-moment inference (e.g., trust breakdown) and cumulative measures (e.g., workload over time).

\subsubsection{Psychological Questionnaires}
We have implemented a suite of standardized psychological questionnaires directly within the VR environment  relevant to transportation simulation using an in-simulation interface, namely in-vr questionnaire \cite{safikhani2021influence}. This approach enables participants to report on their subjective experience without breaking immersion, thereby preserving ecological validity and reducing memory bias. The selected instruments are grounded in validated tools from psychology, cognitive science, social sciences, and environmental research, and are grouped across several conceptual domains. These psychological instruments are synchronized with physiological and behavioral data streams to provide a multimodal understanding of participant state across different modes, environments, and transitions within the simulation. These instruments cover the following domains:
\begin{itemize}
  \item \textbf{Momentary Stress}\\
        As an important aspect of traveler experience, stress is measured using both the momentary stress questionnaire \cite{karvounides2016three}, as well as perceived stress scale \cite{scale1983perceived}.
\end{itemize}

\begin{itemize}
  \item \textbf{Affect and Emotion}\\
        Emotional state is assessed using a two-dimensional valence–arousal scale, capturing both the pleasantness and activation level of current emotions \cite{mauss2009measures} as a momentary assessment as well as using the Positive and Negative Affect Schedule (PANAS) \cite{watson1988development}, which is used to evaluate distinct emotional states such as enthusiasm, irritability, calmness, or anxiety throughout different phases of the simulation.
\end{itemize}

\begin{itemize}
  \item \textbf{Task Demand and Cognitive Load}\\
        Mental workload is measured using the NASA Task Load Index (NASA-TLX) \cite{hart2006nasa}, which includes six subscales: mental demand, physical demand, temporal demand, performance, effort, and frustration. Participants rate each dimension on a continuous scale from 0 (very low) to 100 (very high), allowing for a multidimensional assessment of perceived workload. In scenarios where time or immersion constraints limit full administration, the mental demand and effort subscales are prioritized as reliable proxies for cognitive load within dynamic simulation environments.
\end{itemize}

\begin{itemize}
  \item \textbf{N-back Task Test}\\
        The N-back task is implemented as a secondary task within the simulation to objectively assess cognitive load in dual task conditions as well as working memory \cite{jaeggi2010concurrent}. In this working memory task, participants are required to monitor a sequence of stimuli and identify when the current item matches one presented 'n' steps earlier in the sequence. Higher N-values (e.g., 2-back or 3-back) impose greater demands on working memory and attentional control, providing a reliable behavioral measure of cognitive load. Reaction time, accuracy, and omission errors are recorded and analyzed as indicators of cognitive strain during complex or multitasking segments of the simulation.
\end{itemize}

\begin{itemize}
  \item \textbf{Time Perception}\\
        The perceived passage of time is an important psychological dimension of the travel experience, particularly in public transit and waiting scenarios. We use a validated time perception scale from \cite{altaf2023time} to assess how users subjectively experience the duration of different travel segments. Participants report their perceived duration after completing key phases—such as waiting at a station, riding a crowded bus, or walking through a scenic corridor.
\end{itemize}

\subsubsection{Behavioral Metrics}
Behavioral metrics capture how participants interact with the environment and respond to other agents in the simulation. These data provide insight into decision-making, performance, situational awareness, and interaction quality. The following metrics are collected in real time for each transportation mode, as detailed below:

The Driver Module includes lane-keeping accuracy, defined as deviation from the centerline and frequency of lane departures; braking behavior, measured by reaction time, peak braking force, and smoothness of deceleration; steering input, referring to angular velocity and correction patterns during turns or avoidance maneuvers; gap acceptance, captured as time headway and distance when merging or turning in traffic; and collision avoidance, assessed through time-to-collision, evasive steering, or braking in response to hazards.

The Automated Vehicle (AV) Module includes supervisory gaze patterns, referring to fixation on the road versus the environment during AV operation; takeover behavior, measured by time-to-intervention, control input after takeover, and hesitation delay; override frequency, indicating the number of manual interventions during AV operation; and trust-related behavior, assessed through attention allocation, intervention timing, and hand-off compliance.

The Pedestrian Module includes walking speed, which captures changes in pace under different environmental conditions; path trajectory, measured as deviation from the optimal walking path, hesitation, or rerouting; crossing decision, which considers gap acceptance time, crossing delay, and aborting mid-crossing; and head orientation, described by turn angles and the frequency of visual scanning in intersection zones.

The Public Transit Module includes boarding and alighting behavior, referring to the time it takes to board or exit and any hesitation before entering the vehicle; seat selection behavior, which looks at proximity to others and preference for window or aisle seating; dwell time and movement, including time spent standing or navigating through cabins; and crowd navigation, assessed by route efficiency, avoidance of others, and stop-start frequency in crowded stations.

The Cyclist Module includes pedal cadence, defined as revolutions per minute used to assess rhythm and effort; steering behavior, captured through lateral stability, correction frequency, and swerve avoidance; speed variability, reflecting acceleration and deceleration patterns in response to dynamic road conditions; obstacle response, measured by distance to object at avoidance, brake initiation, and path deviation; and yielding behavior, which considers the cyclist’s reactions to crossing pedestrians or oncoming vehicles.

These behavioral metrics are synchronized with physiological and psychological data to enable multimodal analysis of user states, transitions, and responses in complex transportation environments.

\subsection{Scenario Development Module}
All simulation scenarios were developed using the Unity game engine, which provides a flexible and extensible framework for real-time, multi-agent interaction and immersive environment rendering. To accelerate development and ensure modularity, we utilized a collection of Unity-compatible tools and custom scripts for environment generation, agent behavior programming, animation control, and scene transitions. These tools enabled streamlined implementation of complex interactions, such as pedestrian crossing behaviors, automated vehicle logic, and multimodal transit integration. The scenarios are designed to be fully customizable, allowing researchers to easily modify environment layouts, traffic dynamics, and agent behavior parameters to fit their experimental needs. Each scenario is constructed using a modular hierarchy of prefabs and scripting templates, which can be reused or extended across studies.

\begin{figure}[htbp]
    \centering
    \includegraphics[width=1\textwidth]{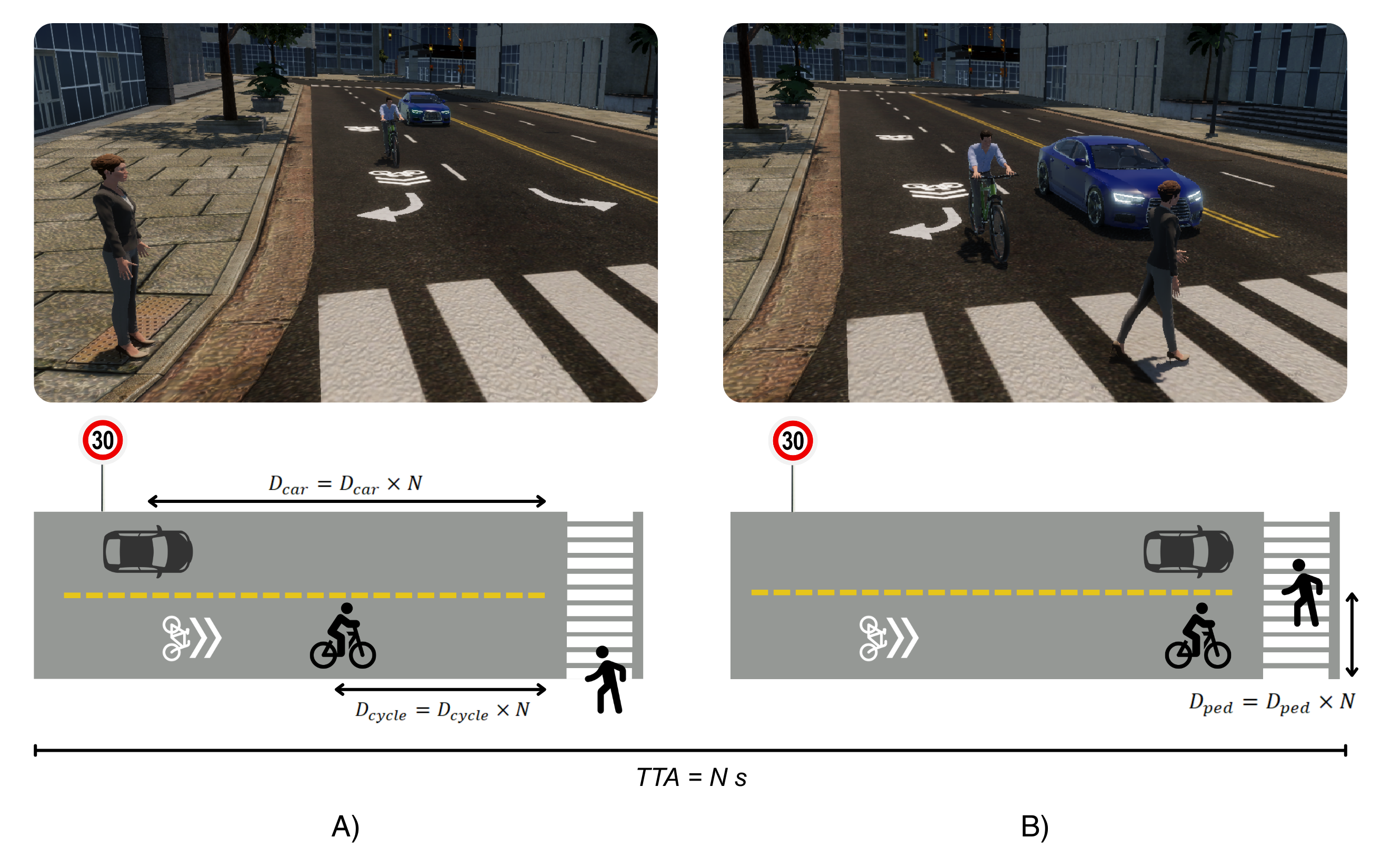}
    \captionsetup{justification=centering}
    \caption{Synchronized Arrival in a Three-Agent Unsignalized Crossing Scenario. A) Initial setup: Agents positioned so their time-to-arrival (TTA) is equal. B) Interaction: All agents arrive at the crosswalk together for negotiation.}
    \label{fig:Synchronized-Arrival-Three-Agent-Unsignalized-Crossing-Scenario}
\end{figure}

The scenario development module also allows control over several scenario-specific variables, including the initial positions, speeds, and time-to-arrival (TTA) of each agent. To create a multimodal interaction scenario at a specific location, such as an intersection or crosswalk, agents can be positioned so that their TTAs to the crossing point are approximately equal—as shown in Figure ~\ref{fig:Synchronized-Arrival-Three-Agent-Unsignalized-Crossing-Scenario}. This configuration increases the chance that agents will arrive at the location around the same time, enabling naturalistic observation of yielding behavior, hesitation, or conflict resolution among road users.

Figure~\ref{fig:Synchronized-Arrival-Three-Agent-Unsignalized-Crossing-Scenario} illustrates the configuration of an unsignalized intersection scenario. Panel A shows the initial positions of the car, cyclist, and pedestrian, each placed at a distance from the crosswalk so that their TTA is identical at the start of the scenario. The car is assigned a speed of 30 km/h. The cyclist speed of 15 km/h is based on field measurements of average cycling speed on shared pedestrian-cycle paths, where observed values in mixed-use environments ranged from 14–16 km/h, with 15 km/h typical near pedestrian crossings \cite{eriksson2019cyclists}. The pedestrian speed of 1.5 m/s is chosen to represent average adult walking speed as found in large-scale field studies of urban crosswalks\cite{knoblauch1996field}. Panel B depicts the critical moment where all three agents arrive at the crosswalk simultaneously, enabling direct interaction and negotiation of right-of-way.

\section{Use Cases}

\begin{figure}[htbp]
    \centering
    \includegraphics[width=1\textwidth]{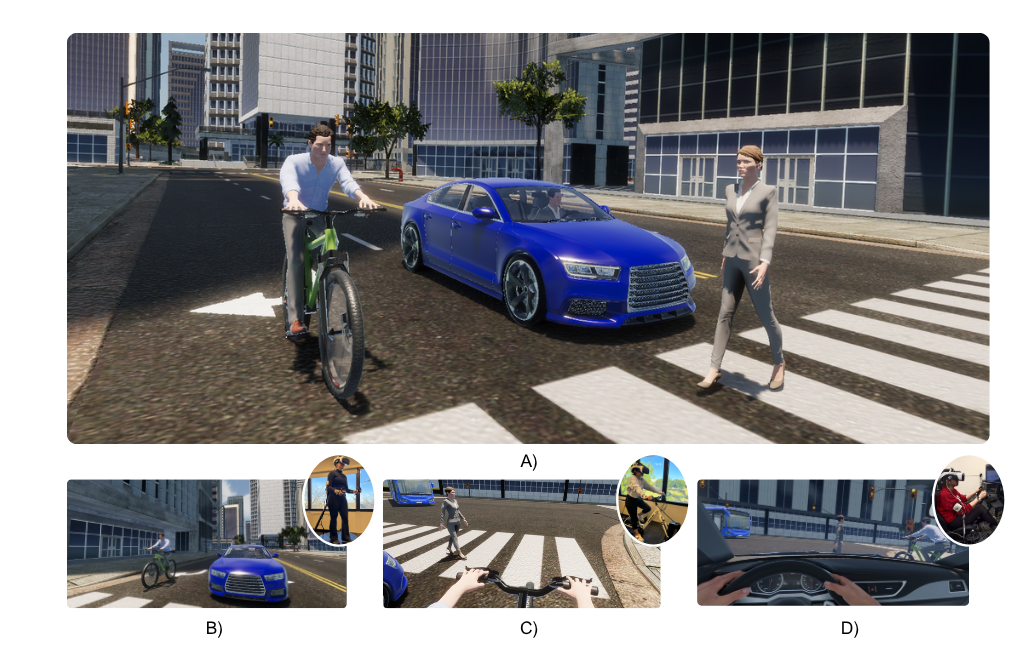}
    \captionsetup{justification=centering}
    \caption{Representative Use Case: A) Shared urban intersection scenario B) Cyclist, C) Pedestrian, and D) Driver views— first-person VR perspective with the external participant view.}
    \label{fig:Multi-Agent-Intersection}
\end{figure}

While virtual reality platforms have been previously used in transportation research, our goal is to highlight the unique capabilities and research opportunities enabled by a human-centered, multi-agent simulation system such as the one presented in this paper. By supporting real-time interactions among diverse transportation users and integrating multimodal sensing, this platform opens up new avenues for investigating complex questions related to behavior, trust, well-being, and system design. A representative example of our multi-agent system is shown in Figure~\ref{fig:Multi-Agent-Intersection}, which depicts a shared urban intersection involving a pedestrian, a cyclist, and an automated vehicle. The figure also includes first-person VR perspectives for each role, illustrating how participants can experience the environment from different viewpoints within the same scenario. Below, we present several representative use cases that further illustrate the breadth and flexibility of the system.

\subsection{Use Case 1: Multimodal Travel Transitions}

The first use case explores the dynamics of multimodal travel, specifically focusing on how users transition between transportation modes—such as cycling to walking, and then boarding public transit—within a single continuous journey. These transitions are increasingly common in urban travel and have important implications for accessibility, usability, safety, and traveler well-being, yet remain underexplored in simulation studies that often isolate individual modes. The simulation platform supports such transitions by allowing a single participant to switch between agent roles, with each segment synchronized to behavioral, physiological, and psychological sensing. This enables analysis of how modal shifts influence user experience and cognitive load across different travel contexts.

\begin{figure}[t]
    \centering
    \includegraphics[width=\textwidth]{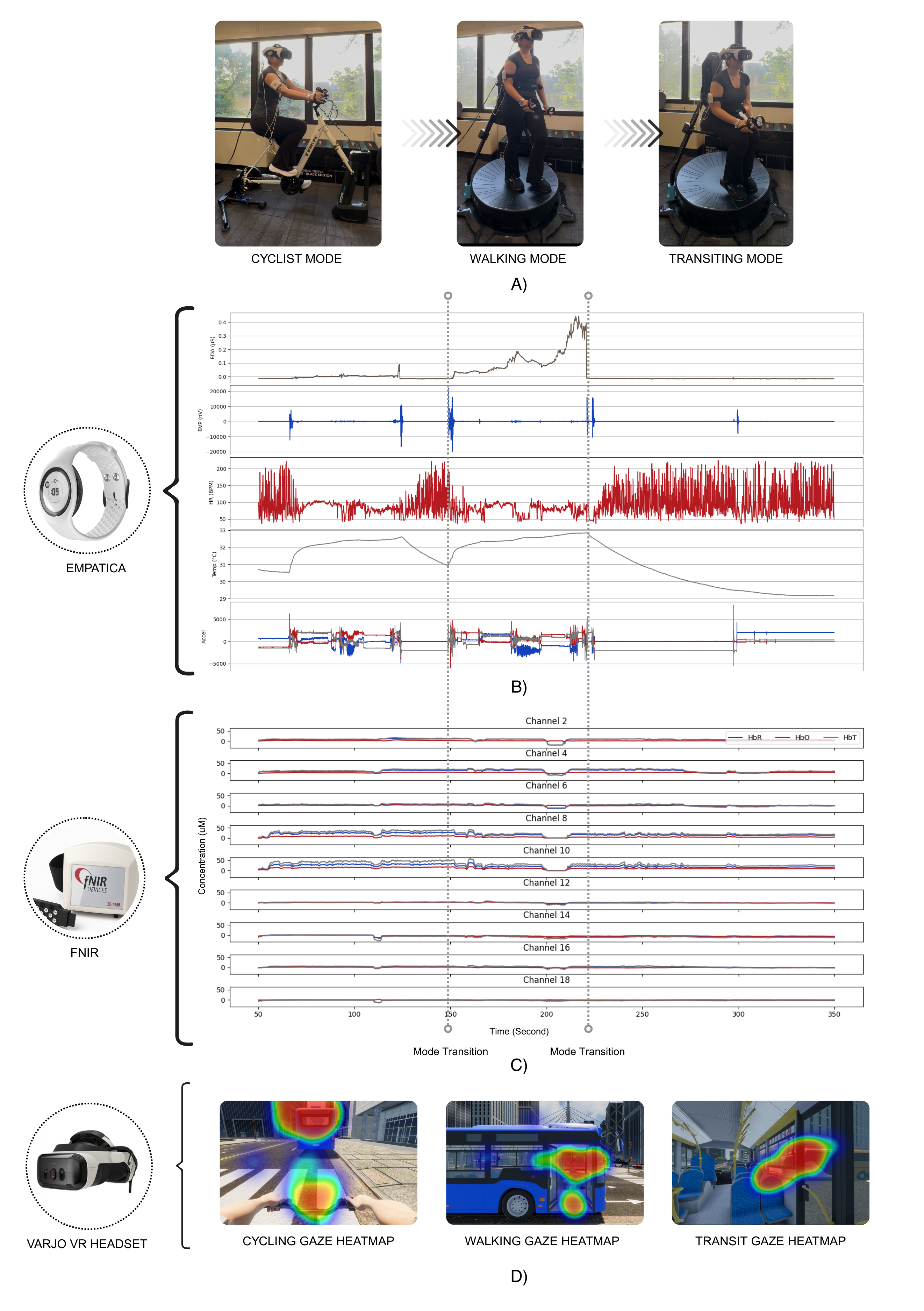}
    \captionsetup{justification=centering}
    \caption{Multimodal Travel Transitions: Sample Data Streams from a Single Participant During Cycling, Walking, and Riding Public Transit}
    \label{fig:Multimodal-Travel-Transitions}
\end{figure}

More specifically, figure~\ref{fig:Multimodal-Travel-Transitions} presents a representative multimodal travel session in which a participant sequentially transitions through cycling, walking, and transit modes using the simulation platform. These transitions are visualized alongside synchronized streams of physiological, neural, and visual attention data collected through embedded and wearable sensing systems. Panel (A) shows the participant physically engaging in each mode, illustrating the seamless progression across transportation types enabled by the system. The second panel (B) displays physiological signals captured by the Empatica EmbracePlus device, including electrodermal activity (EDA), blood volume pulse (BVP), heart rate (HR), skin temperature, and accelerometer data—providing continuous insight into arousal, cardiovascular activity, thermoregulation, and physical movement. Panel (C) illustrates brain activity collected through a headband-style fNIRS sensor, with time-synchronized traces across multiple channels showing changes in oxygenated (HbO), deoxygenated (HbR), and total hemoglobin (HbT) concentrations as indicators of cognitive workload. Finally, panel (D) presents gaze heatmaps derived from the Varjo XR-4 headset’s integrated eye tracker, visualizing spatial attention patterns across the cycling, walking, and transit contexts. These heatmaps reflect differences in visual exploration strategies and attentional demands across modes, offering additional insight into how users perceive and navigate multimodal travel scenarios. Together, the layered data streams highlight the simulation platform’s ability to capture nuanced psychophysiological responses during real-time mode transitions.

\subsection{Use Case 2: Human-Automated Vehicle Interactions}

The second use case focuses on the study of interactions between humans and automated vehicles (AVs), a topic of growing importance as autonomous systems become more prevalent on public roadways. Traditional AV studies often rely on scripted interactions or focus solely on the behavior of the AV itself, limiting our understanding of how real humans—whether drivers, pedestrians, cyclists, or transit users—perceive and respond to AV behavior in complex, dynamic environments.

\begin{figure}[htbp]
    \centering
    \includegraphics[width=1\textwidth]{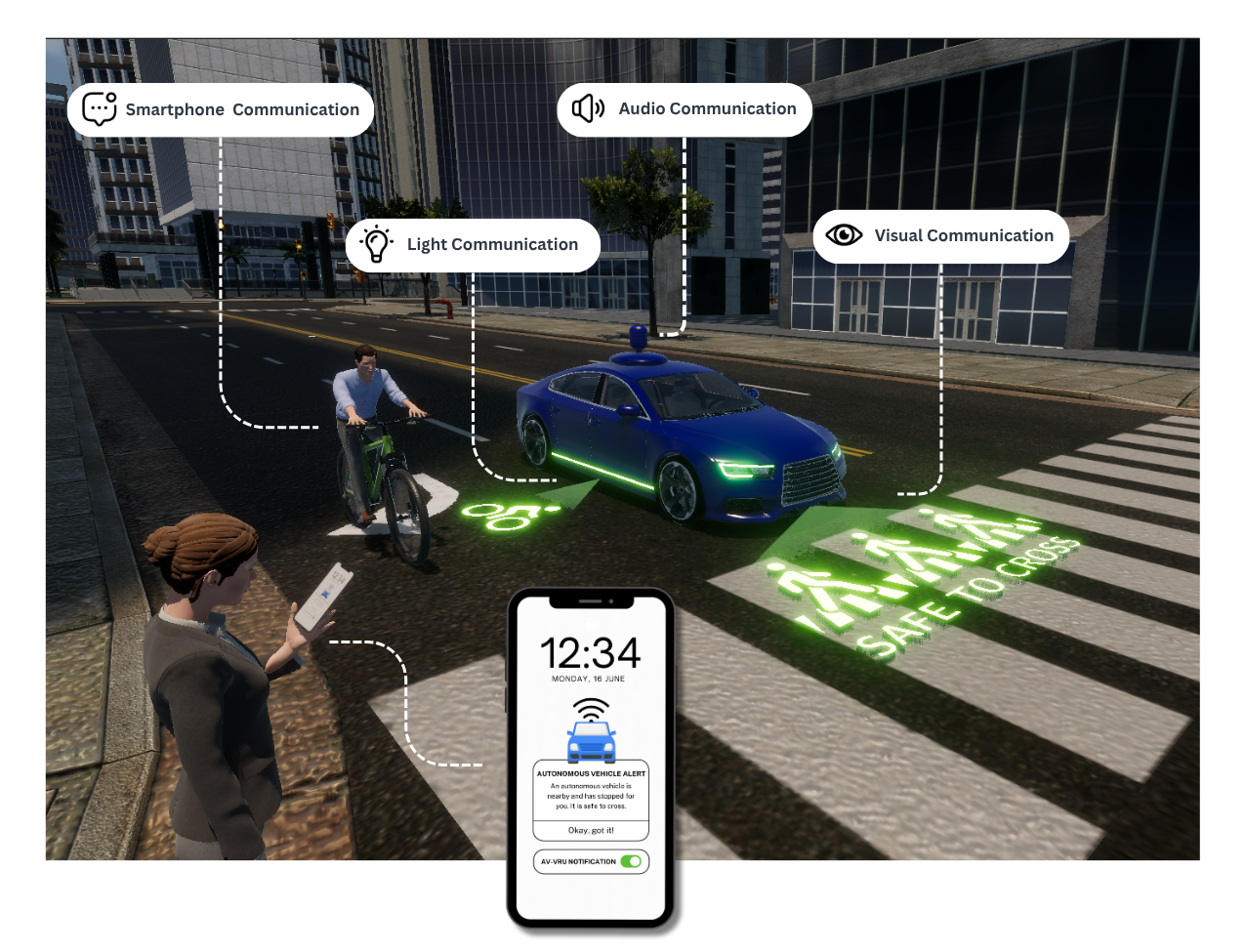}
    \captionsetup{justification=centering}
    \caption{Implemented Multimodal External Human-Machine Interfaces (eHMIs) for AV–VRU Communication in Simulation}
    \label{fig:AV-VRU-eHMI}
\end{figure}

Our multi-agent simulation environment enables real-time, human-in-the-loop interaction between AVs and multiple users within the same virtual space. This setup makes it possible to develop AVs that are not only safe but are socially aware of other road users especially vulnerable road users such as cyclist and pedestrians. Figure~\ref{fig:AV-VRU-eHMI} illustrates a scenario from our simulation platform in which a cyclist and a pedestrian encounter an autonomous vehicle equipped with multiple external human-machine interfaces (eHMIs). These eHMIs span four communication modalities: ground-level visual projections, vehicle-mounted light signals indicating the AV’s awareness and yielding intent, auditory communication delivered through external vehicle speakers, and smartphone-based alerts that notify pedestrians of the AV’s presence and yielding behavior. 

\clearpage
\thispagestyle{empty} 
\clearpage

Within these interactions, participants' psychophysiological responses are continuously monitored using the human sensing module, including fNIRS, Empatica EmbracePlus, and eye tracking. This enables real-time assessment of attentional focus, arousal, trust, and decision-making under varying communication designs. By combining eHMI variation with synchronized sensing data, this use case supports the evaluation of such interaction both objectively and subjectively.

\subsection{Use Case 3: Road User Well-Being}

The third use case focuses on assessing the well-being of road users in response to diverse transportation environments and interactions. With integrated psychological and physiological sensing capabilities, the platform enables a nuanced understanding of how contextual factors—such as roadway design, traffic density, infrastructure elements, and mode choice—affect road user experience beyond traditional performance metrics.

Using this system, researchers can examine multiple dimensions of well-being, including stress, affective states, physical exertion, cognitive workload, and perceived safety. This system enables the study of cross-modal dynamics, such as the effect of aggressive driving on pedestrian stress. The dimensions of well-being are quantified through both subjective and objective metrics, including self-reported psychological questionnaires, physiological signals such as heart rate, electrodermal activity, and skin temperature, and behavioral indicators like gaze patterns, movement variability, and route choices.

\begin{figure}[t]
    \centering
    \includegraphics[width=\textwidth]{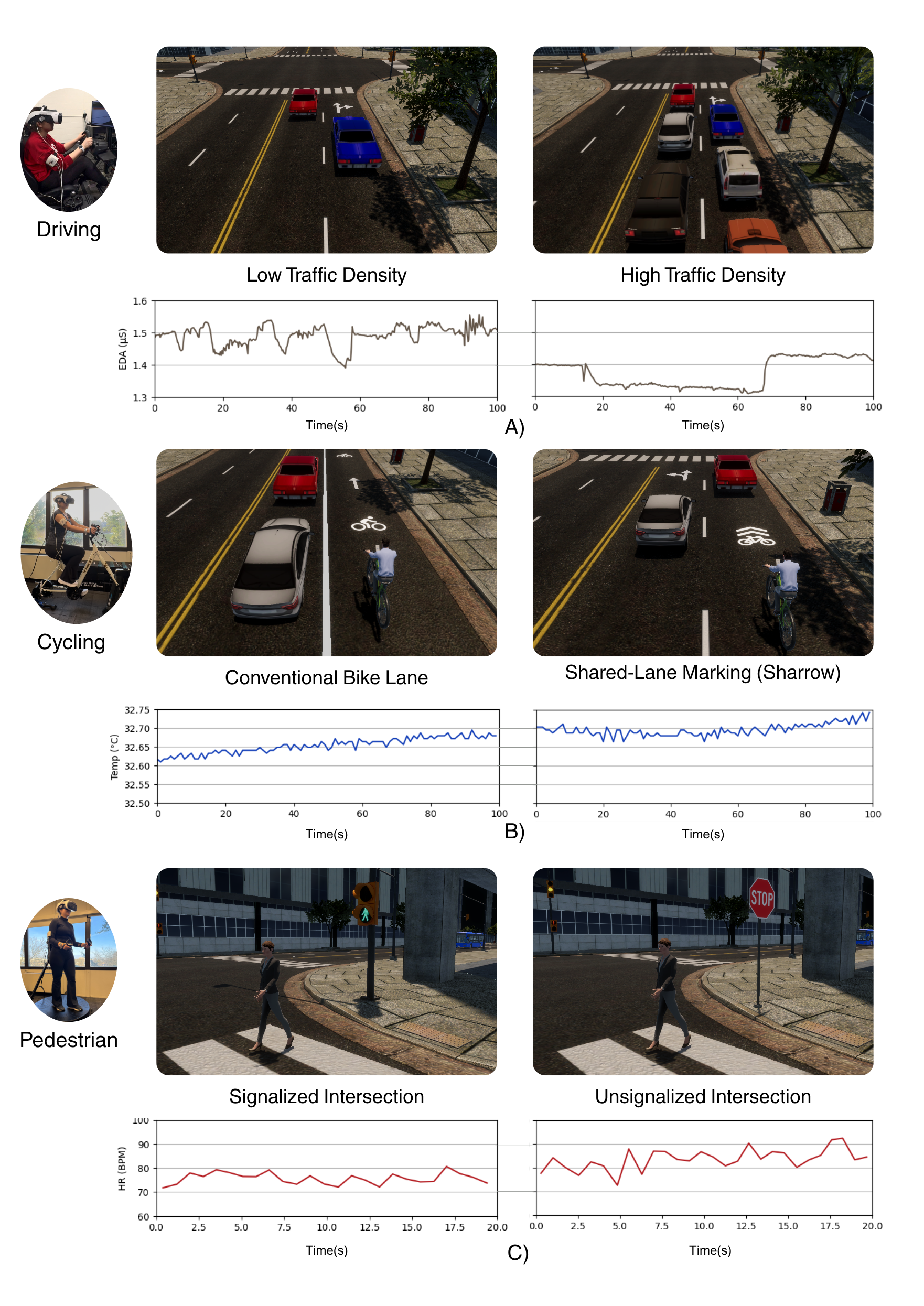}
    \captionsetup{justification=centering}
    \caption{ Sample Mode-Specific Physiological Responses to Contextual Variations in Simulated Transportation Environments:
A) Driving – Electrodermal activity (EDA) under low vs. high traffic density conditions; B) Cycling – Skin temperature under shared-lane marking vs. conventional bike lane conditions; C) Pedestrian – Heart rate (HR) under signalized vs. unsignalized intersection conditions.}
    \label{fig:Contextual-Factors-Physiological-Responses}
\end{figure}

Figure \ref{fig:Contextual-Factors-Physiological-Responses} illustrates selected examples of mode-specific physiological responses to contextual factors that impact user well-being within the simulation.Panel (A) depicts a driving scenario in which electrodermal activity (EDA) is compared between low and high traffic density conditions, reflecting changes in physiological arousal under varying flow environments. Panel (B) presents a cycling scenario comparing skin temperature when riding in a shared-lane marking (sharrow) condition versus a conventional painted bike lane, highlighting thermal and physiological differences due to the presence or absence of designated cycling infrastructure. Panel (C) illustrates a pedestrian scenario involving a signalized versus unsignalized intersection, where heart rate (HR) is used to assess the cardiovascular response to crossing conditions with and without active control.

These representative cases demonstrate how specific design elements and environmental conditions elicit measurable changes in psychophysiological state. Beyond the scenarios shown in Figure \ref{fig:Contextual-Factors-Physiological-Responses}, the simulation platform allows researchers to adjust contextual parameters across several categories. Traffic conditions can be modified in terms of traffic density and modal composition. Control mechanisms include elements such as signal timing strategies and speed limits. Infrastructure design features involve crosswalk layout and bicycle facility type. This configurability supports controlled comparisons of how specific environmental and operational changes influence user well-being across different transportation modes.

\clearpage
\thispagestyle{empty} 
\clearpage

\section{Discussion}

This paper provides a detailed description of a novel approach to developing multi-agent transportation simulation systems. While building on prior literature, our platform addresses key limitations by enhancing behavioral sensing, integrating multimodal human state sensing, and increasing the number of agent types present within a single, synchronized environment. By incorporating pedestrians, cyclists, drivers, automated vehicles, and public transit users into a shared, real-time virtual space, the system allows researchers to examine the interconnected nature of road user behavior in ways that traditional single-agent simulators cannot. These interactions reveal important dynamics such as behavioral dependencies, social signaling, and conflict resolution across modes—factors that are increasingly recognized as critical in human-centered transportation research \cite{sabeti2023mad,lindner2022coupled}.

We depicted three major use cases for this system—multimodal travel transitions, human–automated vehicle interactions, and road user well-being, highlighting the breadth and versatility of the platform. These scenarios were intentionally selected to reflect pressing and highly active areas of research in the transportation literature, such as effect of various transportation modes on users \cite{ettema2016travel,avila2018effects,singleton2019walking}, trust and communication with AVs \cite{zhang2020expectations,gao2021trust,walker2023trust}, and user stress and perception in response to infrastructure design \cite{tavakoli2022driver,tavakoli2022multimodal,guo2023psycho,guo2024unveiling,rodriguez2022level}. While our goal was not to demonstrate statistical significance, these case studies effectively illustrate the applicability of the platform and the sensitivity of its multimodal sensing measures. By capturing detailed physiological, psychological, and behavioral responses in real time, the system provides researchers with a powerful tool to explore nuanced user experiences. Such a platform can substantially advance current research efforts in these domains and enable new inquiries that span across disciplines and user groups.

As the field moves toward more human-centered approaches—supported by recent work using physiological and cognitive sensing in transportation simulators—the role of integrated human sensing has become essential for understanding how individuals perceive, experience, and respond to transportation systems \cite{guo2023psycho,tavakoli2022driver,tavakoli2022multimodal,goodridge2024gaze,boboc2024leveraging}. In this platform, we move beyond traditional behavioral sensing to incorporate deep human sensing across neurological, physiological, and psychological domains. This integration is not merely additive; it enables researchers to investigate potential causal mechanisms underlying behavior, rather than relying solely on correlational analyses. Many recent studies have emphasized the value of human sensing in revealing the mechanisms that drive perception, decision-making, and social interaction within and outside of transportation contexts \cite{tremblay2017social,guo2023psycho,martins2024neural,hopko2024brain}. In other words, by directly measuring internal states such as stress, attentional focus, and cognitive workload, researchers can better explain why a user behaves a certain way in response to a given traffic context—rather than simply observing that the behavior occurred.

One key important aspect of this system is the inclusion of public transit users in human centered simulation systems, which are often neglected in multi-agent simulation studies. This inclusion allows for more realistic modeling of multimodal journeys and shared spaces \cite{sadeghi2023affective,henriquez2025modelling,adjorlu2024virtual}, addressing a critical gap in simulation research, where public transit users are often excluded despite their real-world importance \cite{adjorlu2024virtual,burova2023virtual,sadeghi2023affective}. This is particularly important for studying issues such as accessibility, crowding, time perception, and user comfort—especially as transit environments are often where diverse user groups intersect and interact. Capturing the nuanced behavior of seated, walking, or waiting transit users fills a critical gap in current simulation frameworks and opens the door to new research on urban design and public space optimization.

While multi-agent simulators have only recently begun to gain traction in transportation research, we hope this article fosters an open forum that encourages broader engagement across disciplines and institutions. This platform is intentionally designed with openness and community collaboration at its core. By releasing the architecture, design specifications, and software components as open-source resources, we invite researchers and educators from diverse backgrounds to replicate, extend, and enhance this work. Community-driven development will be critical in validating experimental protocols, expanding use cases, and maintaining relevance as technologies and societal needs continue to evolve. Through this shared effort, we aim to accelerate the adoption of human-centered, multi-agent simulation as a standard approach in transportation research.

While this system represents a significant advancement in enabling multi-agent, human-centered transportation simulation with real users across four distinct modes, several limitations remain. Most importantly, the physical infrastructure required—including high-end VR headsets, motion platforms, and omnidirectional treadmills—may limit the system's replicability for institutions with constrained budgets or space. Although we provide detailed implementation guidance, the cost and footprint of the system components can pose adoption challenges. While our architecture supports real-time synchronization across multiple agents, the current implementation is designed for a local network environment. Extending the platform to support over-the-network simulation could address scalability and collaboration needs, but it also introduces new challenges—most notably latency and synchronization drift—which can affect the fidelity of agent interactions and user experience \cite{sabeti2023mad}. 

Another challenge relates to user comfort. Compared to traditional simulation platforms, immersive VR systems are more likely to induce motion sickness or discomfort, particularly during prolonged sessions or when high-motion scenarios are involved \cite{cossio2025cybersickness,smyth2019motion,talsma2023validation}. While we added various design features such as using high refresh rate headsets to minimize motion sickness, individual sensitivity to VR environments can still vary. This has been widely reported in simulator validation and VR fidelity studies, where motion sickness remains a persistent issue \cite{stoffregen2017effects,cho2022ridevr,chang2021effects,chang2020virtual}. To further support sensory realism and reduce discomfort, we incorporated physical feedback mechanisms in key modules. For instance, following along the prior work in this arena \cite{guo2023psycho}, in the cyclist module, the Wahoo Headwind fan delivers wind resistance scaled to virtual speed, providing congruent airflow that aligns with visual motion. In the driving module, the Next Level Motion Plus actuator simulates vehicle acceleration, braking, and turning forces in real time. These additions help reduce sensory mismatch—a primary cause of VR-induced motion sickness—by providing the body with tactile and vestibular cues that mirror what is seen in the virtual environment. While these strategies improve immersion and comfort, motion sensitivity remains participant-dependent and may still require additional mitigation approaches in future deployments.

Another challenge relates to the combined placement of head-mounted displays and human sensing equipment. Although our human sensing module integrates multimodal data streams—including physiological, neurological, and behavioral measures—certain modalities, such as functional near-infrared spectroscopy (fNIRS), can pose integration difficulties when used alongside VR headsets. These challenges primarily stem from motion artifacts, sensor displacement, and physical interference between the fNIRS headband and the headset hardware \cite{spencer2018exploring, brigadoi2014motion, ocklenburg2023monitoring}. In particular, achieving long-term or continuous monitoring with high spatial and temporal precision remains technically demanding, especially during motion-intensive scenarios or extended sessions. Further advancements in lightweight, low-profile sensor design and artifact correction methods are needed to enable seamless integration in fully immersive simulations.

While our framework is built on a flexible Unity-based virtual environment, the realism of urban scenes and the behavior of AI-controlled agents may not fully capture the complexity and variability of real-world traffic contexts. This limitation is consistent with prior studies that highlight the challenges of modeling naturalistic behavior in AI agents \cite{guo2024unveiling,manawadu2015analysis}. Extending scenario design to incorporate context-specific behavioral scripts and procedurally generated urban environments remains an open challenge. However, we anticipate that as this area of research continues to grow, the availability of open-source models and behavioral datasets will increase—facilitating more realistic, adaptive, and scalable simulation environments as the research community collectively advances this work.

To expand the capabilities, accessibility, and impact of the platform, several future directions are proposed. The most important path forward is to explore alternative, lower-cost hardware configurations that maintain core functionality while reducing the financial and spatial barriers to adoption. This includes modular deployment strategies using mobile VR headsets, compact motion platforms, and scalable sensing options that can be tailored to individual research needs. Additionally, extending the system’s architecture to support remote and cloud-based simulation will enable distributed experimentation across institutions, similar to prior work in this arena \cite{sabeti2023mad}. This extension requires the development of secure, low-latency networking protocols that can ensure real-time agent synchronization and reliable data sharing, even in geographically dispersed environments with limited or unstable internet connectivity.

Moreover, we will continue to refine the integration of human sensing technologies—particularly by exploring new hardware that minimizes physical interference with VR headsets (e.g., low-profile fNIRS bands), and by developing hybrid sensing pipelines that combine physiological, behavioral, and contextual data to improve inference accuracy and robustness. Additionally, improving environmental realism within the simulation will be an ongoing focus. We plan to integrate procedural urban scene generation, intelligent AI agent behaviors, and dynamic contextual elements (e.g., lighting, weather, and crowds) to support more ecologically valid experiments. 

Lastly, now that the platform and associated tools have been open-sourced, we encourage the broader research community to participate in its validation and extension. Community-driven studies, replications, and refinements will be essential to establishing external validity, identifying edge cases, and extending the platform to new domains and user populations. By fostering collaborative development, we hope to accelerate innovation and standardization in human-centered transportation simulation.

\section{Conclusion}
This paper presented a novel, human-centered, multi-agent simulation platform designed to study transportation behavior across pedestrians, cyclists, drivers, automated vehicles, and public transit users. By integrating real-time physiological, behavioral, and psychological sensing within an extensible VR environment, the system enables high-fidelity, interactive studies of complex, multimodal mobility scenarios. We introduced a modular, cookbook-style framework that lowers technical barriers and supports widespread adoption by researchers and educators. Through detailed use cases, we demonstrated the platform’s potential for studying user well-being, human-vehicle interaction, and multimodal travel transitions. This work lays the foundation for more inclusive, scalable, and ecologically valid transportation simulation research.

\section{Acknowledgment}
This material is based upon work supported by the National Science Foundation under Grant No. \#2347012. Any opinions, findings, and conclusions or recommendations expressed in this material are those of the author(s) and do not necessarily reflect the views of the National Science Foundation. The authors would like to also thank the Villanova University College of Engineering for their support in the development of the multimodal simulator. This work received funding from Villanova University’s Falvey Library Scholarship Open Access Reserve (SOAR) Fund.


\bibliographystyle{plain}  
\bibliography{cas-refs}

\appendix
\section{Hardware Components}
\label{appendix:hardware}

List of hardware components used to build the simulation for driving, cycling, pedestrian, and public transit modules. The components are detailed in Table~\ref{tab:hardware}.
\label{appendix:hardware}

\begin{table}[htbp]
  \centering
  \renewcommand{\arraystretch}{1.2}
  \begin{tabular}{|p{3.5cm}|p{3.5cm}|p{4cm}|p{1.5cm}|}
    \hline
    \multicolumn{1}{|c|}{\textbf{Component}} & 
    \multicolumn{1}{c|}{\textbf{Model / Type}} & 
    \multicolumn{1}{c|}{\textbf{Description}} & 
    \multicolumn{1}{c|}{\textbf{Quantity}} \\
    \hline
    VR Headset and Controllers &
      Varjo XR-4 &
      {\small VR headset with integrated eye tracking and handheld controllers for natural interaction and hand motion tracking.} &
      3 \\
    \hline
    Driving Simulator Cockpit &
      GTTrack Simulator Cockpit &
      {\small Full Vehicle chassis with adjustable seating and controls} &
      1 \\
    \hline
    Motion Platform &
      Next Level Motion Plus &
      {\small Provides force feedback simulating vehicle dynamics} &
      1 \\
    \hline
    Racing Wheel \& Pedals &
      Logitech G923 &
      {\small Steering wheel, pedals, and shifter with force feedback} &
      1 \\
    \hline
    Omnidirectional Treadmill &
      Kat Walk VR Core 2+ &
      {\small Allows 360-degree walking freedom and real-time movement tracking} &
      1 \\
    \hline
    Cycling Trainer &
      Wahoo KICKR Trainer &
      {\small Smart trainer providing real-time resistance and cycling performance data} &
      1 \\
    \hline
    Elevation Simulator &
      Wahoo Climb &
      {\small Adjusts bike tilt to simulate realistic road gradients} &
      1 \\
    \hline
    Wind Resistance Fan &
      Wahoo Headwind &
      {\small Provides dynamic wind resistance scaled to cyclist speed and terrain} &
      1 \\
    \hline  
    fNIRS Headband &
      fNIR2000C &
      {\small Measures changes in cerebral blood oxygenation to monitor cognitive workload and brain activity} &
      1 \\
    \hline
    Empatica Wristband &
      Empatica EmbracePlus &
      {\small Monitors physiological signals including heart rate variability and skin conductance} &
      3 \\
    \hline
  \end{tabular}
  \caption{Hardware components required to replicate or extend the system.}
  \label{tab:hardware}
\end{table}



\end{document}